\documentclass[12pt]{article}
\usepackage{amsfonts}
\usepackage{amsmath}
\newcommand{\be}{\begin{equation}}
\newcommand{\ee}{\end{equation}}
\newcommand{\bea}{\begin{eqnarray}}
\newcommand{\eea}{\end{eqnarray}}
\def \la{\label}

\def\({\left (}
\def\){\right )}
\def\]{\right]}
\def\<{\left <}
\def\>{\right>}
\newcommand{\br}{\mathbf{r}}
\newcommand{\bu}{\mathbf{u}}
\newcommand{\bx}{\mathbf{x}}
\newcommand{\bk}{\mathbf{k}}

\newcommand{\A}{\mathbf{A}}
\newcommand{\bp}{\mathbf{p}}
\def \b\xi {{\pmb{\xi}}}
\newcommand{\bxi}{\pmb{\xi}}
\newcommand{\ba}{\pmb{\alpha}}

\newcommand{\F}{\mathcal{F}}
\newcommand{\D}{\mathrm{D}}
\newcommand{\bX}{\mathbf{X}}

\newcommand{\cl}{\mathcal{L}}

\newcommand{\bmu}{\pmb{\mu}}
\newcommand{\bsigma}{\pmb{\sigma}}

\newcommand{\e}{\mathrm{e}}
\renewcommand{\r}{{\mathbf{r}}}
\renewcommand{\k}{{\mathbf{k}}}
\renewcommand{\d}{\mathrm{d}}

\newcommand{\Tr}{\mathrm{Tr}}

\newcommand{\LL}{{\cal L}}

\def\bsl{\boldsymbol}

\begin{document}

\begin{center}

{\Large\bf Thermal quantum electrodynamics of non relativistic charged fluids}

\vspace{5mm}

Pascal R. Buenzli\footnote{Supported by the Swiss National Foundation for Scientific Research}
$\!\!^,$\footnote{E-mail address: pbuenzli@dfi.uchile.cl}, Philippe A. Martin\footnote{E-mail address:Philippe-Andre.Martin@epfl.ch} {\normalsize and} 
Marc D. Ryser
\\ Institute
    of Theoretical Physics\\ Swiss Federal Institute of Technology
    Lausanne\\ CH-1015, Lausanne EPFL, Switzerland\\

\end{center}

\vspace{5mm}

\abstract{The theory relevant to the study of matter in equilibrium with the radiation field is thermal quantum electrodynamics (TQED). We present a formulation of the theory, suitable for non relativistic fluids, based on a joint functional integral representation of matter and field variables. In this formalism cluster expansion techniques of classical statistical mechanics become operative.
They provide an alternative to the usual Feynman diagrammatics in many-body problems which is not perturbative with respect to the coupling constant. As an application we show that the effective Coulomb interaction between quantum charges is partially screened by thermalized photons at large distances. 
More precisely one observes an exact cancellation of the dipolar electric part of the interaction, so that the asymptotic particle density correlation is now determined by relativistic effects. It has still the $r^{-6}$ decay typical for quantum charges, but with an amplitude strongly reduced by a relativistic factor.}

\section{Introduction}

A precise and complete description of equilibrium states of non relativistic quantum charges interacting via the static Coulomb potential has been thoroughly developed in recent years in the low density regime \cite{AlCoPe}-\cite{BaMa}. This description relies on the use of the Feynman-Kac path integral representation of the thermal Gibbs weight allowing for a classical-like analysis of thermodynamic potentials and particle correlations. Essentially, quantum point charges are mapped onto 
a set of closed Brownian paths (loops) whose random shapes account for the quantum fluctuations. Techniques of classical statistical mechanics become available in the auxiliary phase space of loops, in particular the method of cluster expansion (Mayer graphs). The latter is particularly suited to calculations in dilute systems, where the small parameter is the density. 

Low density expansions of the pressure are performed up to the order $\rho^{5/3}$ \cite{AlCoPe}, exact asymptotics of particle correlations are determined in \cite{Cornu1}.
Phases with atomic or molecular recombination can also be conveniently studied, e.g.  the equation of state \cite{AlBaCoMa} and  the van der Waals forces \cite{AlCoMa} in the Saha regime, as well as the dielectric response of an atomic gas \cite{BaMa} 
(see \cite{Al}, \cite{BrMa} for reviews and additional references). 
However, none of these works take into account the coupling of the charges to the radiation field which is responsible for both 
effective magnetic interactions (Lorentz forces) and retardation effects. The purpose of this paper is to show how the above formalism and techniques can be generalized when matter is thermalized with the quantized electromagnetic field. It is an extension of \cite{ElBBuMa} (hereafter referred to as I) where the field was considered as classical. When the field is quantized in the transverse gauge, it is appropriate to represent the Gibbs weight by means of the bosonic functional integral based on the coherent state representation of photon states. In this way the quantum field is mapped onto a set of classical-like random electromagnetic fields with (imaginary) time dependent amplitudes. Since the energy of the free field is quadratic in the field amplitudes, the latter are distributed with Gaussian statistics.
At this stage, quantum charges can, as in  \cite{AlCoPe}-\cite{BaMa}, be put into correspondence with Brownian charged loops with the aid of the Feynman-Kac-It\^o formula. The coupling to the field appears as the flux of the magnetic field accross the loops.
Thus TQED becomes isomorphic to a system of random charged wires (the loops) experiencing a random magnetic field. The calculation rules are entirely defined by the covariances of the processes associated to the loops and to the field amplitudes,   
together with the use of Wick's theorem.
In this setting, the cluster (Mayer or virial) expansions of classical statistical mechanics can again be put at work providing an alternative to the standard TQED Feynman graph calculations, which is not perturbative with respect to the coupling constant (namely, the electric charge). The method is particularly adapted to study equilibrium phases of plasmas and recombination processes in presence of the electromagnetic field at moderate density .

In Section 2 we describe the actual system consisting of non relativistic charges interacting with the photon field. In order to make sense mathematically and physically, the model requires a high energy cut-off defined by $\hbar \omega_{ \bk_{{\rm cut}}}=\bar{m} c^2$ to eliminate photons that are more energetic than the rest mass energy of a particle of typical mass $\bar{m}$ ($\hbar$ is the Planck constant, $c$ the speed of light and $\omega_\bk=ck$ the photon frequency for the wave number $\bk,\;k=|\bk|$). This gives a typical wave number cut-off $k_{{\rm cut}}=\bar{m}c/\hbar$ with corresponding wave length $\lambda_{{\rm cut}}
=\hbar/\bar{m}c$ (see e.g. \cite{Cohen}, Chap. 3, for a discussion of this point).
High energy processes, such as pair creation or annihilation, demand for the use of the relativistic wave equation (Klein-Gordon or Dirac). They are not taken into account in this model whose predictions only make therefore sense for distances 
$r\gg \lambda_{{\rm cut}}$. 

The construction of the relevant functional representations are recalled in Section 3 for the field and in Section 4 for the particles. 
Since the subject is well developed elsewhere we merely present the main structure in a perspective adapted to our purposes (see references in Section 3).
The thermalized photon field involves  the typical energy
$\hbar\omega_{\bk_{\rm ph}}=\hbar c{k_{\rm ph}} =k_BT=\beta^{-1}$, with corresponding wave length $\lambda_{\rm ph}=\beta\hbar c$, called the thermal length of the photon ($T$ is the temperature and $k_B$ the Boltzmann constant). On the other hand, the mean kinetic energy of a nonrelativistic particle $\epsilon_
{\bk_{{\rm mat}}}=(\hbar k_{{\rm mat}})^2/2\bar{m}=k_BT$ defines the de Broglie thermal wave length of the particle $\lambda_{{\rm mat}}=\hbar\sqrt{\beta/\bar{m}}$.
To be consistent with non relativistic particle motion  we must impose that the thermal energy imparted to the particle in the form of kinetic energy is much lower than its rest mass energy, namely $\epsilon_{\bk_{{\rm mat}}}=k_B T\ll  \bar{m}c^2$,
implying
\be
\lambda_{{\rm cut}}=\frac{\lambda_{{\rm mat}}}{\sqrt{\beta \bar{m}c^2}}\ll\lambda_{{\rm mat}}\ll\lambda_{\rm ph}=\sqrt{\beta \bar{m}c^2}\lambda_{{\rm mat}}
\la{8.1}
\ee
where $\beta \bar{m}c^2\gg 1$ is a dimensionless relativistic parameter.
Therefore, when the field is quantized, we have to distinguish two different regimes at large distance $r$ 
\be
  \lambda_{{\rm mat}}\ll\lambda_{{\rm ph}}\ll r \label{8.1a}
\ee
or
\be
\lambda_{{\rm mat}}\ll r\ll \lambda_{{\rm ph}}\;\;.
\label{8.1b}
\ee
In addition to the quantum lengths, there are typical classical lengths such as the interparticle distance $a=\rho^{-1/3}$ ($\rho$ the density) and the Debye screening length $\lambda_D$. The latter do not enter explicitly in our subsequent analysis because the regime (\ref{8.1a}) of main interest in this paper deals with distances $r$ far beyond $\lambda_{{\rm mat}},\; a$ and $ \lambda_D$. 
 For instance, in an electrolyte the lengths $\lambda_{{\rm mat}},\; a$ and $ \lambda_D$ are of the same order of magnitude ($\sim 10^{-10}-10^{-9}$ m) but they are all much smaller than $\lambda_{{\rm ph}}\sim10^{-5}$ m (see concluding remarks). 
We shall only require that the density is low enough for the system to be in a fluid phase so that we can apply the standard methods of statistical mechanics (cluster expansions).

In Section 5, we determine the effective potential between loops
arising when the field degrees of freedom have been integrated out. This can easily be done by
a  Gaussian integration, as in paper I. Indeed, a simple structure shows up from the fact that in the functional integral representation the coupling of matter to the field amplitudes occurs
linearly in a phase factor (in contrast to the original quantum Hamiltonian which has a coupling quadratic in the creation and annihilation operators). Then the whole effect of the field is contained in an effective potential depending on $\lambda_{{\rm mat}}$ and $\lambda_{\rm ph}$ that can be viewed as a current-current interaction between pairs of loops (Formula (\ref{7.18}) in Section 5). 

We use these results in Section 6 to find the behaviour of the particle correlations in both regimes (\ref{8.1a}) and (\ref{8.1b}). 
Equipped with the Coulomb potential and this new
effective field-induced potential, all standard rules of classical statistical mechanics can be applied to the calculation of particle correlations (some care has to be exercised with the computation rules for stochastic integrals, see appendix A). It is seen that the large distance behaviour of the correlation is determined by the square of dipoles fluctuations, the total dipole of a loop having a part due to its charge and a part due to its current. 
This leads to a generic $r^{-6}$ decay of the correlation.
Now a striking phenomenon occurs in case (\ref{8.1a}) above: namely the screening of the dominant part of the Coulomb interaction by thermalized photons. When $r\gg\lambda_{{\rm ph}}$, the transverse field has a contribution that 
 exactly cancels the dipolar electric part of the loop fluctuations. Only current fluctuations of the loops are left,  
which cannot be screened. In this regime, the correlation still has a 
$r^{-6}$ decay, but with a relativistic prefactor $(\beta \bar{m}c^2)^{-2}$.

In paper I, we have argued that large distances are controlled by small wave numbers of the radiation field and the latter can therefore be treated classically. This apparently sensible argument proves to be incorrect in the sense that it does not predict the aforesaid Coulombic cancellation which results of a subtle conspiracy between the Planck constants of field and matter. It might be inconsistent, in the transverse gauge, to make a classical approximation for the radiation part of the field only. Approximations should be made in a fully gauge invariant manner. Note however that, once the cancellation has been taken into account, the theory of paper I
correctly predicts the remaining correlation tail induced by the current fluctuations.

In the regime (\ref{8.1b}), the radiation field has essentially no incidence on the decay of the particle correlations and one recovers the purely Coulombic tail due to electrical dipole flucuations as the dominant contribution, plus terms vanishing as $r/\lambda_{{\rm ph}}\to 0$. More generally, all results of  \cite{AlCoPe}-\cite{BaMa}
are expected to remain valid in this regime up to tiny relativistic corrections.

Other applications for which the present formalism will be relevant are suggested in the concluding remarks.

\section{The model}
The non relativistic QED model consists of non relativistic quantum charges (electrons, nuclei, ions) with
masses $m_{\gamma}$ and charges $e_{\gamma}$. They obey the appropriate Bose or Fermi statistics and interact with the quantum electromagnetic field, the latter being relativistic by nature. The index $\gamma$ labels the ${\cal S}$ different species and runs from $1$ to ${\cal S}$. The particles are confined in a box $\Lambda\in\mathbb{R}^3$ of linear size $L$ whereas  the field itself is enclosed in a large box $K$ with sides of length $R,\,R\gg L$. The Hamiltonian of the total finite volume system reads in Gaussian units
\begin{align}
    H_{L,R}=\sum_{i=1}^{N}\frac{\({\bf
            p}_{i}-\frac{e_{\gamma_{i}}}{c}\A(\r_{i})\)^{2}}{2m_{\gamma_{i}}}+
    \sum_{i<j}^{N}
    \frac{e_{\gamma_{i}}e_{\gamma_{j}}}{|\r_{i}-\r_{j}|}+\sum_{i=1}^{N} V_{{\rm
            ext}}(\gamma_{i},\r_{i})+H_{0}^{{\rm rad}} \la{B.1}\;\;.
\end{align}
The sums run on all particles with position $\r_{i}$, momentum $\mathbf{p}_i$ and species index
$\gamma_{i}$,\;i=1,\ldots,N\;, $V_{{\rm ext}}(\gamma_{i},\r_{i})$ comprises a possible external potential plus a steep wall potential
that confines the particles in $\Lambda$. The latter can eventually be taken infinitely
steep at the wall's position implying Dirichlet boundary conditions
on the particle wave functions at the boundaries of $\Lambda$.  

The electromagnetic field is written in the Coulomb (or transverse) gauge so that the vector potential $\A(\r)$ is divergence free and $H_{0}^{{\rm rad}}$ is the Hamiltonian of the free radiation field. 
We impose periodic boundary conditions on the faces of the large box $K$. Expanding $\A(\r)$ and the free photon energy
$H_{0}^{{\rm rad}}$ in the plane wave modes $\bk= (\frac{2\pi
    n_{x}}{R},\frac{2\pi n_{y}}{R},\frac{2\pi n_{z}}{R})$ gives
\begin{align}
    \A(\r)&=\(\frac{4\pi \hbar c^{2}}{R^{3}}\)^{1/2}\sum_{\bk\lambda}g(k)
    \frac{ {\bf e}_{\bk\lambda}}{\sqrt{2\omega_{\bk}}}
    (a_{\bk\lambda}^{\dagger}\e^{-i\bk\cdot\r}+a_{\bk\lambda}\e^{i\bk\cdot\r})\la{B.2}\\
    H_{0}^{{\rm rad}}&={\sum_{\bk\lambda}}\hbar
    \omega_\bk\,a_{\bk\lambda}^{\dagger}a_{\bk\lambda} \la{B.2a}\;,
\end{align}
where $a_{\bk\lambda}^{\dagger}, \,a_{\bk\lambda}$ are the creation and annihilation
operators for photons in the mode $({\bk\lambda})$ with commutation relations $[a_{\bk\lambda},\; a_{\bk'\lambda'}^{\dagger}]=\delta_{\lambda\lambda'}\delta_{\bk\bk'}$, ${\bf e}_{\bk\lambda}$
($\lambda=1,2$) are two unit polarization vectors orthogonal to ${\bk}$ and
$\omega_\bk=ck,\; k=|\bk|$.  In (\ref{B.2}) $g(k)$ is a
real spherically symmetric smooth form factor that takes
care of the ultraviolet divergencies. It obeys $g(0)=1$ and is supposed to decay rapidly beyond the characteristic wave number $k_{\rm cut}=\bar{m} c/\hbar$. Note that in (\ref{B.1}) we have  included neither the Pauli coupling $-\bmu\cdot {\bf B}(\r)$ of the electronic spin with the magnetic field ${\bf B}(\r)=\nabla\wedge  \A(\r)$ ($\bmu=(e\hbar/4m_ec )\bsigma$ is the magnetic moment of the electron, $\bsigma$ are the Pauli matrices) nor the nuclear hyperfine interaction (see comments in the concluding remarks). 
It is known that the Hamiltonian (\ref{B.1})
is $H$-stable \cite{BuFrGr} for a finite ultraviolet cutoff $k_c^{-1}<\infty$, namely $H_{L,R}$ possesses an extensive lower bound proportional to the total number of particles (for a review of $H$-stability in non relativistic QED, see \cite{Gr}).

We are interested in the situation in which matter and photons are in thermal equilibrium at the same temperature $T$. 
The total partition function associated with (\ref{B.1}) 
\begin{align}
    Z_{L,R}= \Tr\ \e^{-\beta H_{L,R}}
    \la{B.3}
\end{align}
is obtained by carrying out the trace $\Tr=\Tr_\text{mat} \Tr_\text{rad}$ of the total Gibbs
weight over particles' and the field's degrees of freedom, namely over the particle wave functions with appropriate quantum statistics and the Fock states of the photons.
The corresponding free energy density in the thermodynamic limit will be defined by 
extending to infinity first the field region $K$ 
 and then the box $|\Lambda|$ containing the charges. Thus the excess free energy relative to that of the free radiation field is
\be
f= -k_BT\lim_{L\to\infty}\frac{1}{|\Lambda|}\lim_{R\to\infty}(\ln Z_{L,R}-\ln Z_{0,R}^{{\rm rad}})\;,
\label{B.3a}
\ee
where $Z_{0, R}^\text{rad}=\Tr_\text{rad}\exp{(-\beta H_{0}^{{\rm rad}})}$ is the partition function of the free field.
A lower bound for $f$ has been established in \cite{LiLo}, but at the moment, to our knowledge, a complete proof of the existence of the thermodynamic limit has not yet been provided. Nevertheless we shall assume that the quantities of interest in this paper have a well-behaved thermodynamic limit.

As in I, we shall be concerned in the sequel with the partial average \begin{align}
   [\e^{-\beta H_{L,R}}]_{\text{mat}} =\frac{\Tr_\text{rad} \e^{-\beta H_{L,R}}}{Z^\text{rad}_{0, R}}
    \la{B.4}
\end{align}
giving the (non normalized) statistical distribution of matter obtained by averaging on the degrees of freedom of the radiation field. The corresponding normalized reduced density matrix is
\footnote{Here the notation is slightly different from paper I where $\rho_{L,R}$ (I.5) designates the field averaged quantity (\ref{B.4}). }
\be
\rho_{L,R}=\frac{\Tr_\text{rad} \e^{-\beta H_{L,R}}}{Z_{L, R}}
=\frac{  [\e^{-\beta H_{L,R}}]_{\text{mat}}}{\Tr_\text{mat} [\e^{-\beta H_{L,R}}]_{\text{mat}}}\;\;.
\la{B.4a}
\ee

It will be convenient to single out in $H_{L,R}$ the free radiation part writing 
\begin{align}
   & H_{R,L}=H_\mathbf{A}+H_{0}^{{\rm rad}},\label{tot}\\&
     H_\mathbf{A}= \sum_{i=1}^N \frac{\({\bf
            p}_{i}-\frac{e_{\gamma_{i}}}{c}\A(\r_{i})\)^{2}}{2m_{\gamma_{i}}}
    +U_\text{pot}(\r_{1},\gamma_{1},\ldots,\r_{N},\gamma_{N}) 
    \la{B.4b}
\end{align}
where
\be
U_\text{pot}(\r_{1},\gamma_{1},\ldots,\r_{N},\gamma_{N})= 
 \sum_{i<j}^{N}
    \frac{e_{\gamma_{i}}e_{\gamma_{j}}}{|\r_{i}-\r_{j}|}+\sum_{i=1}^{N} V_{{\rm
            ext}}(\gamma_{i},\r_{i})
\la{B.4c}            
\ee
is the total potential energy.

\section{Functional integral representation of the field}

If the field is treated classically (namely the creation and annihilation operators are replaced by c-number amplitudes) it is immediately seen that the free field distribution factorizes in the total Gibbs weight as $\exp\({-\beta H_{R,L}}\)=\exp\({-\beta H_{0}^{{\rm rad}}}\)\exp\({-\beta H_\mathbf{A}}\)$.
Thus the partial trace (\ref{B.4}) reduces to integrals with a Gaussian weight since the free radiation part $\exp\({-\beta H_{0}^{{\rm rad}}}\)$ is Gaussian in the field amplitudes, a fact that was exploited in I. 

If the field is quantized, it is first of all necessary to represent the electromagnetic field by c-functions in the total Gibbs weight $\exp\({-\beta H_{R,L}}\)$. This can be achieved by means of 
the standard functional integral for bosonic quantum field \cite{NeOr}, \cite{Papa}. We briefly recall its construction.
First, one considers the coherent states associated to the field modes 
\begin{align}
    |\alpha_{\bk\lambda}\rangle=\sum_{m=0}^{\infty}\frac{\big(\alpha_{\bk\lambda}
        a^{\dagger}_{\bk\lambda}\big)^{m}}{m!}|0\rangle=\e^{\alpha_{\bk\lambda}
        a^{\dagger}_{\bk\lambda}}|0\rangle,\quad
    a_{\bk\lambda}|\alpha_{\bk\lambda}\rangle=\alpha_{\bk\lambda}|\alpha_{\bk\lambda}\rangle
    \la{7.1}
\end{align}
They have scalar products 
\be
\langle\alpha_{\bk\lambda}|\alpha_{\bk'\lambda'}\rangle
=\e^{\alpha_{\bk\lambda}^\ast\alpha_{\bk' \lambda'}'} 
\label{7.1a}
\ee
and the closure relation reads
\begin{align}
    \int\! \frac{\d^{2}\alpha_{\bk\lambda}}{\pi}\ \e^{-|\alpha_{\bk\lambda}|^2}
    |\alpha_{\bk\lambda}\rangle \langle\alpha_{\bk\lambda}|=\mathbf{1}\;\;. 
 \la{7.2}
\end{align}
We denote $\ba =\{\alpha_{\bk\lambda}\}_{\bk\lambda}$,
$|\ba\rangle=\prod_{\bk\lambda}|\alpha_{\bk\lambda}\rangle$,
$d\ba=\prod_{\bk\lambda}\frac{d^{2}\alpha_{\bk\lambda}}{\pi}$, $\ba
\ba'$=\\$\sum_{\bk\lambda} \alpha_{\bk\lambda}\alpha_{\bk\lambda}^{'}$, etc., and
introduce the infinite product representation \\$\e^{-\beta
    H}=\lim_{M\rightarrow\infty}\left(\mathbf{1}-\frac{\beta}{M}H\right)^M$ where $H\equiv H_{L,R}=H_\mathbf{A}+ H_{0}^{{\rm rad}} $ is the total Hamiltonian operator (\ref{tot}). 
 Using this representation and inserting
$M-1$ closure relations one can write the following coherent state matrix element as 
\begin{align}
    &\langle\ba |\e^{-\beta
        H}|\ba\rangle=\lim_{M\rightarrow\infty}\left[\prod_{l=1}^{M-1}\int\!\!
        d\ba_{l}\ \e^{-\ba_{l}^{*}\ba_{l}}\right]\la{7.3}\\
    &\times
    \langle\ba|\left(\mathbf{1}-\tfrac{\beta}{M}H\right)|\ba_{M-1}\rangle\cdots
    \langle\ba_{l}|\left(\mathbf{1}-\tfrac{\beta}{M}H\right)|\ba_{l-1} \rangle
    \cdots\langle\ba_{1}|\left(\mathbf{1}-\tfrac{\beta}{M}H\right)|\ba
    \rangle\nonumber\;.
\end{align}
As a first step we consider the partial coherent state matrix element
$\langle \ba_{l}| \e^{-\beta H} | \ba_{l-1}\rangle$, which is still an operator acting on the Hilbert space of the particle states. 
Its evaluation is achieved by putting $H$ in normal order. Using (\ref{7.1a}), this yields
\begin{align}
    \langle\ba_{l}|\left(\mathbf{1}-\tfrac{\beta}{M}H\right)|\ba_{l-1}\rangle =
    \e^{\ba_{l}^{*}\ba_{l-1}}(1-\tfrac{\beta}{M}H(\ba_{l}^*,\ba_{l-1})),
\la{7.4}
\end{align}
where $H(\ba_{l}^*,\ba_{l-1})$ depends on the complex amplitudes $\ba$ according to the normal order form of $H$.
From (\ref{tot}), (\ref{B.4b}), one finds
\begin{align}
&H(\ba_{l}^*,\ba_{l-1})=H_{\A}(\ba_{l}^*,\ba_{l-1})+ D_N+  H_{0}^{{\rm rad}}(\ba_{l}^*,\ba_{l-1})\nonumber\\
&H_{\A}(\ba_{l}^*,\ba_{l-1})=  \sum_{i=1}^{N}\left[\frac{\({\bf p}_{i} - \frac{e_{\gamma_{i}}}{c}
            \A(\r_{i}, \ba_{l}^*,\ba_{l-1})\)^{2}}{2m_{\gamma_{i}}}
         \right]  + U_\text{pot}(\r_{1},\gamma_{1},\ldots,\r_{N},\gamma_{N})
\la{7.5}
\end{align}
where the vector potential $\A(\r_{i},\ba_{l}^*,\ba_{l-1})$ has the same form as
in (\ref{B.2}) with the operators $a^{\dagger}_{\bk\lambda},a_{\bk\lambda}$ replaced by the
complex amplitudes $\alpha_{l,\bk\lambda}^*,\alpha_{l-1,\bk\lambda}$, and likewise for $H_0^\text{rad}(\ba_l^*,\ba_{l-1})$. 
The constant
\begin{align}
D_N=\sum_{i=1}^Nd_{\gamma_i},\quad d_{\gamma_{i}} = 
 \frac{2\pi\hbar}{c}\frac{e_{\gamma_{i}}^2}{m_{\gamma_{i}}}\(\frac{1}{R^3}
    \sum_{\bk}\frac{g^2(\bk)}{k}\) 
    \la{7.5a}
\end{align}
arises when putting $\(\A(\r_{i})\)^{2}$ in normal order.  Inserting (\ref{7.4})
in (\ref{7.3}) yields
\begin{align}
    &\langle\ba |\e^{-\beta
        H}|\ba\rangle=\lim_{M\rightarrow\infty}\left[\prod_{l=1}^{M-1}\int\!\!
        d\ba_{l} e^{-\ba_{l}^{*}(\ba_{l}-\ba_{l-1})}\right]\la{7.6}
   \\&\times \left(1-\tfrac{\beta}{M}H(\ba^*,\ba_{M-1})\right)
    \cdots\left(1-\tfrac{\beta}{M} H(\ba_{l}^*,\ba_{l-1})\right) \cdots
   \left(1-\tfrac{\beta}{M}H(\ba_{1}^*,\ba)\right) \;\;.\nonumber
\end{align}
One introduces the formal functional integral as usual by interpreting
$\alpha_{l, \bk\lambda}=\alpha_{\bk\lambda}(\frac{l}{M})$ as the value at
$\tau=\frac{l}{M}$ of a closed trajectory $\alpha_{\bk\lambda}(\tau)$ in the
complex plane, 
$\alpha_{\bk\lambda}(0)=\alpha_{\bk\lambda}(1)=\alpha_{\bk\lambda}$. The parameter $\tau$, $0\leq \tau\leq 1,\;$ is a dimensionless imaginary time.
In the limit $M\to \infty$ the product of infinitesimal evolutions in
(\ref{7.6}) tends to the time ordered propagator
\begin{align}
&\e^{-\beta D_N}{\cal T}\big[\e^{-
    \beta\int_{0}^{1}\d\tau H(\ba^*(\tau+\eta),\ba(\tau))}\big]=
 \nonumber\\& \e^{-\beta D_N} \;\e^{-\beta\int_{0}^{1}\d\tau H_0^\text{rad}(\ba^*(\tau+\eta),\ba(\tau))} {\cal T}\big[\e^{-
    \beta\int_{0}^{1}\d\tau H_\mathbf{A}(\ba^*(\tau+\eta),\ba(\tau))}\big],\quad \eta\to 0^+\;\;.
\la{7.6a}
\end{align}
The imaginary time ordering ${\cal T}$ is necessary because although the field amplitudes $\ba(\tau)$ are now c-functions,  
the $H(\ba^*(\tau+\eta),\ba(\tau))$ are still operators acting
on the space of particle wave functions and therefore they do not commute for different times. However, the free field part $H_0^\text{rad}(\ba^*(\tau+\eta),\ba(\tau))$ commutes with the matter dependent part $H_\mathbf{A}(\ba^*(\tau+\eta),\ba(\tau))$ (\ref{7.5}) and can be factorized out of the ${\cal T}$-product according to the second line of (\ref{7.6a}). The $\eta\to 0^+$ prescription means that, as a result of the normal order, the amplitudes correponding to the creation operators $\ba^*(\tau+\eta)$  have to be evaluated in (\ref{7.6a}) at times infinitesimaly larger than those corresponding to the annihilation operators $\ba(\tau)$ (see (\ref{7.6})).
Finally, (\ref{7.6}) can be written in the condensed form of a path integral
\begin{align}
    \langle\ba |\e^{-\beta H}|\ba\rangle=  \e^{-\beta D_N} \lim_{\eta\to 0_+}&\left[ 
  \int_{\ba(0)=\ba}^{\ba(1)=\ba} \!\! \d[\ba(\cdot)]\,
    \e^{-\int_{0}^{1}\d\tau\big(\ba^{*}(\tau)\tfrac{\partial}{\partial\tau}\ba(\tau)+\beta
        H_{0}^{{\rm rad}}(\ba(\tau))\big)}
        \right. \nonumber \\
        &\times \left.\phantom{\int}\!\!\!\!\;
 {\cal T}\big[\e^{-\beta \int_{0}^{1}\d\tau
        H_\mathbf{A}(\ba(\tau))}\big]\right]_\eta 
        \la{7.7}
\end{align}
where the bracket $[\cdots]_\eta$ indicates that the amplitudes $\ba^*$ in (\ref{7.7})  have to be evaluated at the time $\tau +\eta$. 
The partial Gibbs distribution (\ref{B.4}) is obtained by integrating the matrix element (\ref{7.7}) on $d\ba$ and then dividing it by the partition function of the free field
\be
 [\e^{-\beta H_{L,R}}]_{\text{mat}} =\frac{1}{Z^\text{rad}_{0 R}}  
\int d\ba e^{-|\ba|^2}  \langle\ba |\e^{-\beta H}|\ba\rangle\;\;.
\la{7.7aa}
\ee
More generally, the factor 
\be 
\e^{-\int_{0}^{1}\d\tau\big(\ba^{*}(\tau)\tfrac{\partial}{\partial\tau}\ba(\tau)+\beta
        H_{0}^{{\rm rad}}(\ba(\tau))\big)}
 \la{7.7a}       
\ee 
 in (\ref{7.7})
provides a Gaussian (free) weight on the space of time-dependent complex field amplitudes $\ba(\tau)$. If $F(\ba(\cdot))$ is a functional of these amplitudes, we will denote its average with respect to the distribution  (\ref{7.7a}) by
\begin{align}
&<F(\ba(\cdot))>_\text{rad}=\nonumber\\&\frac{1}{Z^\text{rad}_{ 0R}}
\lim_{\eta\to 0_+}
  \left[ 
  \int \! D\ba\,
    \e^{-\int_{0}^{1}\d\tau\big(\ba^{*}(\tau)\tfrac{\partial}{\partial\tau}\ba(\tau)+\beta
        H_{0}^{{\rm rad}}(\ba(\tau))\big)}F(\ba(\cdot))
        \right]_\eta\;\;.
\la{7.7b}
\end{align}
Here the integral runs over all possible closed paths, setting
$ \int \!  D\ba\cdots=\int d\ba e^{-|\ba|^2} \int_{\ba(0)=\ba}^{\ba(1)=\ba} \!\! \d[\ba(\cdot)]\cdots$. 
It is well known that this Gaussian integral is characterized by the covariance  \cite{NeOr}
\begin{align}
    &\langle\alpha_{{\bk}\lambda}(\tau)\alpha^*_{\bk'\lambda'}(\tau')\rangle_\text{rad}  
    =\delta_{\lambda\lambda'}\delta_{\bk\bk'} \mathcal{C}(k,\tau-\tau')\nonumber \\
    &\langle\alpha_{{\bk}\lambda}(\tau)\alpha_{\bk'\lambda'}(\tau')\rangle_\text{rad}=\langle\alpha^*_{{\bk}\lambda}(\tau)\alpha^*_{\bk'\lambda'}(\tau')\rangle_\text{rad}=0  
 \la{7.10}
 \end{align}
with  
  \begin{align}
\mathcal{C}(k,\tau-\tau')= & e^{-\beta\hbar\omega_{\k}(\tau-\tau')} [\theta(\tau-\tau') (n_{\bk}+1)+\theta(\tau'-\tau) n_{\bk}],\quad \tau\neq\tau' 
\la{7.10z}
 \end{align}
\be
 \mathcal{C}(k,0)=n_{\bk},\quad \tau=\tau'
 \la{7.10y}
\ee
and
\be
n_{\bk}=(\e^{\beta\hbar\omega_{\bk}}-1)^{-1}
\la{7.10b}
\ee
is the Planck distribution ($\theta$ is the Haevyside step function). The function $\mathcal{C}(k,\tau-\tau')$ is discontinuous at $\tau=\tau'$ with the value $\mathcal{C}(k,0)=n_{\bk}$ as a consequence of the normal order prescription $\eta\to 0_+$.  

Functional integrals  (\ref{7.7b}) of the paths $\ba(\tau)$
 are in principle entirely determined by application of Wick's theorem and use of the covariance (\ref{7.10}).  
In particular,  using the representation (\ref{7.7b}), the effective partial  thermal weight (\ref{B.4}) of matter when the field degrees of freedom have been traced out can now be written as
\be
[\e^{-\beta H_{L,R}}]_{\text{mat}} =\e^{-\beta D_N}\;
< {\cal T}\big[\e^{-\beta \int_{0}^{1}\d\tau
        H_\mathbf{A}(\ba(\tau))}\big]>_\text{rad}\;\;.
\la{3.10}
\ee
This will be the starting point of our investigation of the particle correlations in presence of the field in Section 5.

\section{Functional representation of the particles}

We now come to the functional integral representation of the matter degrees of freedom.
One notes that the operator ${\cal T}\big[\e^{-\beta \int_{0}^{1}\d\tau H_\mathbf{A}(\ba(\tau))}\big]$ in (\ref{3.10}) is  
the propagator on the space of particle wave functions associated to the time dependent Hamiltonian $H_\mathbf{A}(\ba(\tau))$  where the vector potential has been replaced by its non-operatorial classical form 
\begin{align}
    \A(\r,\ba(\tau))=\(\frac{4\pi \hbar
        c^{2}}{R^{3}}\)^{1/2}\sum_{\bk\lambda}g(k) \frac{ {\bf
            e}_{\bk\lambda}}{\sqrt{2\omega_{\bk}}}\big(\alpha_{\bk\lambda}^{*}(\tau)
    \e^{-i\bk\cdot\r}+\alpha_{\bk\lambda}(\tau)\e^{i\bk\cdot\r}\big)\;\;.
\la{7.8}
\end{align}
The time dependence is introduced by 
the amplitudes $\alpha_{\bk\lambda}(\tau)$, which are random functions distributed by the Gaussian weight (\ref{7.7a}) of the bosonic functional integral. 
However, for a fixed function \\$\ba(\tau),\,0\leq\tau\leq 1$, $H_\mathbf{A}(\ba(\tau))$ can be viewed as
the Hamiltonian of the particle system submitted to the external vector
potential (\ref{7.8}). In this situation one can apply the Feynman-Kac-It\^o formula \cite{Roepstorff} to represent the configurational matrix element of  ${\cal T}\big[\e^{-\beta \int_{0}^{1}\d\tau H_\mathbf{A}(\ba(\tau))}\big]$. 

 For a single particle of mass $m$ and
charge $e$ in a scalar potential $V_{\text{ext}}(\r)$ and time dependent vector potential
$\A(\r,s)$, we first recall that this matrix element reads \cite{Feynman-Hibbs}, \cite{Roepstorff}, \cite{Simon}
\begin{align}
    &\langle \r |
   {\cal T} \exp\left(\!-\beta\int_0^1 d\tau\left[\frac{\(\bp-\frac{e}{c}\A(\r,\tau)\)^{2}}{2m}+V^\text{ext}(\r)\right]\!\right)|\r\rangle\!=\!
    \(\frac{1}{2\pi\lambda^2}\)^{3/2} \!\!\!\int\!\!\! \D(\b\xi) \nonumber\\
    &\time\tau
    \exp\left(-\beta\left[\int_{0}^{1}\!\!\! \d \tau\ V^\text{ext}\big(\r +\lambda \b\xi
        (\tau)\big)-i\frac{e}{\sqrt{\beta m c^{2}}}\int_{0}^{1} \!\!\! \d\b\xi(\tau) \cdot
        \A\big(\r+\lambda \b\xi (\tau),\tau\big)\right] \right).
        \nonumber \\
    \label{3.1}
\end{align}
Here $\b\xi (\tau),\;0\leq \tau \leq 1,\; \b\xi (0)=\b\xi (1)=\mathbf{0}$, is a closed
dimensionless Brownian path and $\D(\b\xi)$ is the corresponding conditional
Wiener measure normalized to $1$. This measure is Gaussian, formally written as
\be
\D(\b\xi)=\exp\Big(-\frac{1}{2}\int_0^1 \d \tau \left|\frac{\d\b\xi (\tau)}{\d
    \tau}\right|^2\Big)\d[\b\xi (\cdot)]\;\;.
\la{3.1a}    
\ee     
It has zero mean and covariance
\begin{align}
    \int\!\! \D(\b\xi)\,\xi^{\mu}(\tau)\xi^{\nu}(\tau')=\delta_{\mu\nu}(\min(\tau,\:\tau')-\tau \tau')\;,
    \label{3.2}
\end{align}
where $\xi^{\mu}(\tau)$ are the Cartesian coordinates of $\b\xi(\tau)$.  In this
representation a quantum point charge looks like a classical charged closed filament ${\cal F}=(\r,\;\b\xi)$
located at $\r$ and with a random shape $\b\xi(\tau),\;0 \leq\tau\leq 1$,
the latter having a spatial extension given by the thermal de Broglie length $\lambda=\hbar\sqrt{\beta/m}$
(the quantum fluctuation).  The magnetic phase in (\ref{3.1}) is a stochastic line integral: it is the flux of the magnetic field
across the closed filament. The correct interpretation of this stochastic
integral is given by the rule of the middle point, namely, the integral on a small
element of line $\bx-\bx'$ is defined by
\begin{align}
    \int_{\bx'}^{\bx}\!\!\!\d\b\xi\cdot {\bf f}(\b\xi)=(\bx-\bx')\cdot{\bf f}
    \(\frac{\bx+\bx'}{2}\),\quad \bx-\bx'\to 0 \;\;.\la{3.3a}
\end{align}
We shall stick to this rule when performing explicit calculations
\footnote{We find it convenient to apply the middle point rule because it correctly represents the quantum mechanical Gibbs weight in presence of a vector potential (divergenceless or not)   \cite{Roepstorff}. 
Although we shall not use the It\^o prescription we keep the terminology of Feynman-Kac-It\^o formula.}. 
If there is no field, the generalisation of the Feynman-Kac formula  to the many particle system including quantum statistics has been presented in a number of works, see e.g. \cite{Cornu1},
\cite{BrMa}, \cite{Martin}. When the field is present, the analysis presented in the above works can be reproduced without changes, the only difference being the inclusion of the additional phase factor corresponding to the vector potential (see \cite{Cornu2} in the case of a uniform magnetic field). We give here merely the basic formulae resulting from these generalisations.

Filaments ${\cal F}=(\r,\;\b\xi(\tau),\;0\leq \tau \leq 1)$
associated to  single quantum particles
are generalized to Brownian loops
\be
    \LL = (\r, \gamma, q, \bX (\tau)), \qquad 0 \leq \tau \leq q\;\;.
\la{3.3}
\ee
The  $q$-loop  $\LL$ 
consists again in a closed Brownian path, 
\be
   \r (\tau) = \r + \lambda_{\gamma} \bX (\tau), \qquad 0 \leq \tau \leq q,
\la{3.4}
\ee
now parametrised by the (dimensionless) imaginary time $\tau,\;0\leq \tau\leq q$.
The path is specified by its position $\r$ in space, a particle species~$\gamma$, a number of particles~$q$,
and a shape $\bX (\tau)$ with $\bX (0) = \bX (q) = {\bf 0}$. 
The positions of the $q$ particles are located at points 
$\r(k-1)$ on the path, $k=1,...,q$ . The
paths $\bX_r(\tau)$, $r=1,\ldots,n$, corresponding to $n$ different loops are independent random variables 
\be
\langle X_r^\mu(\tau)X_s^\nu(\tau')\rangle_{\bX}=0,\quad r\neq s
\la{indep}
\ee
 and identically distributed according to a normalized
Gaussian measure $D(\bX)$ with covariance
\begin{align}	
&\langle X_r^\mu(\tau)X_s^\nu(\tau')\rangle_{\bX}=   \int D(\bX) X^\mu(\tau)X^\nu(\tau')\nonumber\\
 &= \delta_{\mu\nu}q
\Big[\min\Big(\frac{\tau}{q},\frac{\tau'}{q}\Big)-\frac{\tau\tau' }{q^2}\Big],\quad\quad r=s,\quad\quad\mu,\nu~= 1,2,3\;\;.
\label{cov}
\end{align}
The number $q$ accounts for the quantum statistics of the species $\gamma$, it corresponds to grouping together $q$  particles that are permuted according to a cyclic permutation of length $q$. 
The set of all possible loops (\ref{3.3}) will be called the space of loops. It plays the role of an auxiliary classical-like phase space where methods of classical statistical mechanics can be used. 
Note that for Bose or Fermi quantum statistics, the $N$ particles are distributed into $n$ loops $\LL_r, \;r=1,\ldots,n$, according to their species and $N=\sum_{r=1}^n q_r$.
Maxwell-Boltzmann statistics are recovered if all $q$-loops for $q>2$ are disregarded. Then a loop $\LL$  reduces to a filament $\F$ and the covariance (\ref{cov}) reduces to (\ref{3.2}) so that in this case there is a one-to-one correspondence between filaments and particles.  

The generalisation of the Feynman-Kac-It\^{o} formula to the many-body problem induces loop self-interactions and interactions between loops.
The total energy of a system of $n$ loops has three contributions:
\be
\sum_{r=1}^nU(\LL_r) + U_\text{pot}(\LL_1,\ldots,\LL_n) +
U_\mathbf{A}(\LL_1,\ldots,\LL_n)\;\;.
\la{3.5}
\ee
The potential energy $U_\text{pot}$ of $n$ loops is the sum of pairwise interactions between loops plus the action of external potentials
\be
    U_\text{pot}(\LL_1,\ldots,\LL_n) = \sum_{r<s}^n e_{\gamma_{r}} e_{\gamma_s}
V_c(\LL_r,\LL_s)+\sum_{r=1}^n V_{\rm ext}(\LL_r)
\la{3.6}
\ee
where the interaction between two different loops is Coulombic
\footnote{A local regularization of the Coulomb potential has to be added when dealing with Maxwell-Boltzmann statistics.}
\be	\label{A2 def V(LL_i,LL_j)}
    V_c(\LL,\LL') = \int_0^{q} d \tau \int_0^{q'} d \tau' \,
{\delta}(\tilde{\tau}-\tilde{\tau}')\,
 \frac{1}{| \r(\tau) - \r'(\tau')|}.
\ee
Here, ${\delta}(\tilde{\tau})=\sum_{n=-\infty}^\infty
\delta(\tau-n)$ is the Dirac comb of period one, $\tilde{\tau}=\tau\; {\rm mod}\;1$. Hence
$V_c(\LL_r,\LL_s)$ represents the sum of the  interactions between the
particles in the loop $\LL_r$ and the particles in
the loop $\LL_s$, and the factor ${\delta}(\tilde{\tau}-\tilde{\tau}')$ implements the quantum mechanical constraint of equal time interaction  inherited from the Feynman-Kac-It\^o formula.  

The term $ \sum_{r=1}^nU(\LL_r)$ is the self energy of the loops with
\be
   U(\LL) = \frac{e_\gamma^2}{2} \int_0^qd \tau \int_0^{q} d \tau'\,
(1-\delta_{[\tau],[\tau']})
    	{\delta}(\tilde{\tau}-\tilde{\tau}') \frac{1}{| \r(\tau) - \r'(\tau')|}\;.
\la{3.6a}
\ee
This is the sum of the mutual interactions of the particles within one loop. The
factor $(1-\delta_{[\tau],[\tau']})$, where $[\tau]$ denotes the integer part of $\tau$, avoids counting the proper self-energies of the
point particles; when $q=1$, $U(\LL)$ vanishes.
Finally, 
\begin{align}
    U_\A(\LL_1,...,\LL_n ) =
        -i\sum_{r=1}^{n}\frac{e_{\gamma_{r}}} {\sqrt{\beta
                m_{\gamma_{r}}c^{2}}} \int_{0}^{q_r}\!\!\! \d\bX_r(\tau) \cdot
        \A(\r_r+\lambda_{\gamma_r} \bX_r (\tau), \ba(\tilde{\tau})) 
         \la{3.7}
\end{align}
where $ \ba(\tilde{\tau})$ 
is the periodic extension of $\ba({\tau}),\;0\leq \tau\leq 1$ to all $\tau$. The phase factors in  (\ref{3.7}) 
arise from the interaction of the particles with the vector potential.
They are the flux of the corresponding (periodic) magnetic field across the loops.

The following remark is in order. In (\ref{A2 def V(LL_i,LL_j)})-(\ref{3.7}), $\tau$-integrals run from $0$ to $q$
as a consequence of grouping together in a single path $\bX(\tau),\;0\leq \tau \leq q$, all particles belonging to a permutation cycle of $q$ elements (see \cite{BrMa}, Chap. V, Section A1). Such integrals can as well be reduced to the interval $0\leq \tau \leq 1$ by means of the identity 
\be
\int_0^q d\bX(\tau)F(\bX(\tau),\ba(\tilde{\tau}))=\sum_{m=0}^{q-1}
\int_0^1d\bX(\tau+m)F(\bX(\tau+m),\ba({\tau}))\;\;.
\la{3.7a}
\ee
The notation in  (\ref{A2 def V(LL_i,LL_j)})-(\ref{3.7}) is short and convenient.

The total Gibbs weight on the space of loops (including the normal order constant $D_N$ (\ref{7.5a}))   
\be
\e^{-\beta D_N}\;
\exp\left[-\beta\(\sum_{r=1}^n U(\LL_r)+
U_\text{pot}(\LL_1,\ldots,\LL_n)+U_\A(\LL_1,...,\LL_n \)\right]
\la{3.8}
\ee
gives (up to normalisation) the joint probability distribution of 
n interacting loops in a realisation of the electromagnetic field having amplitudes $\ba(\tau)$. Individual loops have Gaussian weights defined by the covariance (\ref{cov}), thus calculations
of averages on loops reduce in principle to applications of the
Wick theorem. One will also have to consider averages of stochastic integrals involving the line elements $dX^{\mu}(\tau)$. This is achieved by 
supplementing (\ref{indep}) and (\ref{cov}) by the expressions 
 \begin{align}
\langle dX_r^\mu(\tau)X_s^\nu(\tau')\rangle_{\bX}=&
\(\frac{\partial }{\partial \tau}\langle dX_r^\mu(\tau)X_s^\nu(\tau')\rangle_{\bX}\) d\tau,
\nonumber\\ &=\delta_{rs}\delta_{\mu\nu}\(\theta(\tau-\tau')
-\frac{\tau'}{q}\)d\tau,\quad{\rm for}\quad\tau\neq\tau'
\la{3.9}
\end{align}
\begin{align}
\langle d X_r^\mu(\tau)X_s^\nu(\tau)\rangle_{\bX}&=\delta_{rs}\delta_{\mu\nu}\frac{1}{2}\(\frac{d}{d\tau}\langle  X_r^\mu(\tau)X_s^\nu(\tau)\rangle_{\bX} \)d\tau
\nonumber\\ &=\delta_{rs}\delta_{\mu\nu}\(\frac{1}{2}-\frac{\tau}{q}\)d\tau, \quad{\rm for}\quad\tau=\tau'
\la{3.9aa}
\end{align}
and
\begin{align}
\langle d X_r^\mu(\tau)dX_s^\nu(\tau')\rangle_{\bX}=&\(\frac{\partial^2}{\partial \tau\partial \tau'}\langle  X_r^\mu(\tau)X_s^\nu(\tau')\rangle_{\bX} \)d\tau d\tau'
\nonumber\\ &=\delta_{rs}\delta_{\mu\nu}\(\delta(\tau-\tau')-\frac{1}{q}\)d\tau d\tau'\;\;.
\la{3.9a}
\end{align}
These formulae are in accordance with the middle point rule, which assigns
the value $1/2$ to $\theta(\tau -\tau')|_{\tau=\tau' }$ in (\ref{3.9aa}) (see e.g. calculations in the appendix A of I).

At this point we see that computations of thermal properties of the
system of charges and field corresponding to the Hamiltonian 
(\ref{B.1}) are entirely specified by the form of the Gibbs weight (\ref{3.8}) on the space of loops together with the Gaussian distributions of the field amplitudes $\ba(\cdot)$ and loop shapes $\bX_r(\cdot)$. Indeed, the Gibbs weight (\ref{3.8}) is a functional  of $\ba(\cdot)$ and $\bX_r(\cdot)$, and Gaussian averages are uniquely characterized by the covariances
(\ref{7.10z}), (\ref{7.10y}), (\ref{indep}), (\ref{cov}), 
(\ref{3.9}), (\ref{3.9aa}) and  (\ref{3.9a}). 
Of course, calculation rules in the auxiliary space of loops have to be completed by appropriate formulae that relate quantities obtained in the loop formalism to the physical information of interest such as thermodynamic potentials or particle and field correlation. We shall not develop such formulae in general here but will present an application of this formalism to the determination of the particle density correlations in presence of the field in Section 6.   

\section{The effective magnetic potential}

We are now in position to explicitly trace out the field degrees of freedom to obtain the representation of the matter statistical weight $[\e^{-\beta H_{L,R}}]_{\text{mat}}$ (\ref{3.10}) on the space of loops.  The corresponding distribution is obtained by averaging (\ref{3.8}) on the field variables, namely 
\begin{align}
\e^{-\beta D_N}&\;
\exp[-\beta\sum_{r=1}^nU(\LL_r)] 
\exp[-\beta U_\text{pot}(\LL_1,\ldots,\LL_n)] \nonumber\\&\quad
\quad\quad\quad\quad\times\langle\exp[-\beta U_\A(\LL_1,\ldots,\LL_n)]\rangle_\text{rad}
\la{3.10b}
\end{align}
From (\ref{3.7}) and (\ref{7.8}) one sees that  $\exp[-\beta U_\A(\LL_1,\ldots,\LL_n)]$ is a phase factor linear in the field amplitudes $\alpha_{\bk\lambda}(\tau)$ and $\alpha^*_{\bk\lambda}(\tau)$. Since \\$<\cdots>_\text{rad}$ is Gaussian, the average can be performed with the help of the basic formula (written here for a single mode of the field)
\begin{align}
&\left\langle\exp\left[i\int_0^1 d\tau( f(\tau)\alpha^*(\tau)+f^*(\tau)\alpha(\tau))\right]\right\rangle_\text{rad}
=\nonumber \\&
\exp\left[-\int_0^1 d\tau\int_0^1d\tau'f^*(\tau)\langle\alpha(\tau)\alpha^*(\tau')\rangle_\text{rad}f(\tau')\right]\;\;.
\la{3.10a}
\end{align}
To apply this formula we introduce the eigenmode expansion (\ref{7.8}) of the vector potential in (\ref{3.7})
\begin{align}
-\beta U_\A
(\LL_1,\ldots,\LL_n)=
i\left[\sum_{\bk\lambda}
\left(\sum_{r=1}^n\int_0^{q_r}d\bX_r(\tau)\cdot\bu^r_{\bk\lambda}
(\tau)\right)\alpha^*_{\bk\lambda}(\tilde{\tau}) +c.c
\right]
\end{align}
where $\bu^r_{\bk\lambda}(\tau)$ collects the factors
\be
\bu^r_{\bk\lambda}(\tau)=\beta \frac{e_{\gamma_{r}}} {\sqrt{\beta
                m_{\gamma_{r}}c^{2}}}\(\frac{4\pi \hbar
        c^{2}}{R^{3}}\)^{1/2}g(k)
 \frac{ {\bf
            e}_{\bk\lambda}}{\sqrt{2\omega_{\bk}}}
e^{-i\bk\cdot(\br_r+\lambda_{\gamma_r}\bX_r(\tau))}\;\;.
\la{3.11}
\ee
Application of the formula (\ref{3.10a}) gives
\begin{align}
&\langle\exp[-\beta U_\A(\LL_1,\ldots,\LL_n)]\rangle_\text{rad}=
\nonumber\\
&\exp\left[-\sum_{\bk\lambda}\sum_{r=1}^n\int_0^{q_r}d\bX_r(\tau)\cdot(\bu^r_{\bk\lambda}(\tau))^*\;
\sum_{s=1}^n\int_0^{q_s}d\bX_s(\tau')\cdot (\bu^{s}_{\bk\lambda}(\tau'))\;\mathcal{C}(k,\tilde{\tau}-\tilde{\tau}')\right]\;\;.
\la{3.12}
\end{align}
We have used the fact that the covariance (\ref{7.10}) is diagonal with respect to $\bk\lambda$ and $\mathcal{C}(k,\tau-\tau')$ is given by (\ref{7.10z}), (\ref{7.10y}).
The remark made after (\ref{3.7}) applies also here. In order to use (\ref{3.10a}), all $\tau$-integrals can as well be reduced to the interval $0\leq \tau\leq 1$ by means of the formula (\ref{3.7a}). Then $\mathcal{C}(k,\tilde{\tau}-\tilde{\tau}')$ is the periodic continuation of $\mathcal{C}(k,{\tau}-{\tau'}),\; 0\leq \tau,\tau'\leq 1$.  
Since $\bu^r_{-\bk\lambda}(\tau)=\pm(\bu^r_{\bk\lambda}(\tau))^*$, it is clear that by changing $\bk\to -\bk,\;r\to s$ in (\ref{3.12}) only the even part of $\mathcal{C}(k,\tilde{\tau}-\tilde{\tau'})$ contributes.
One finds from (\ref{7.10z}) for $\tau\neq 0$ 
\begin{align}
\mathcal{C}_{\text{even}}(k,{\tau})&=\frac{1}{2}[\mathcal{C}(k,{\tau})+\mathcal{C}(k,{-\tau})]\nonumber\\
&=n_\bk\cosh(\beta \hbar \omega_\bk\tau )+ \frac{1}{2}e^{-\beta \hbar \omega_k|\tau|}\nonumber\\
&=\frac{\cosh[\beta \hbar \omega_\bk(|\tau|-1/2)]}{\sinh(\beta \hbar \omega_\bk/2)}
\la{3.13}
\end{align}
whereas from (\ref{7.10y}) 
\be
\mathcal{C}_{\text{even}}(k,0)= n_\bk,\quad \tau=0\;\;.
\la{3.13a}
\ee
Introducing the explicit form of $\bu^r_{\bk\lambda}(\tau)$ (\ref{3.11}), equation (\ref{3.12}) becomes
\begin{align}
&\langle\exp[-\beta U_\A(\LL_1,\ldots,\LL_n)]\rangle_\text{rad}=\nonumber\\&
\exp\(-\beta\sum_{r,s=1}^n\frac{4\pi \hbar e_{\gamma_r}e_{\gamma_s}}{\sqrt{m_{\gamma_r}m_{\gamma_s}}}\int\frac{d^3\bk}{(2\pi)^3}\frac{g^2(k)}{2\omega_\bk}e^{i(\bk\cdot(\r_r-\r_s)}\delta_{\mu\nu}^{\text tr}(\bk)\right.
\nonumber\\& \times\left.\left[
\int_0^{q_r}dX_r^\mu(\tau)\int_0^{q_s}dX_s^\nu(\tau')
e^{i\bk\cdot(\lambda_{\gamma_r}\bX_r(\tau)-\lambda_{\gamma_s}\bX_s(\tau'))}\mathcal{C}_{\text{even}}(k,\tilde{\tau}-\tilde{\tau}')
\right]\)\;\;.
\la{3.14a}
\end{align}
The transverse delta function $\delta_{\mu\nu}^{\text tr}(\bk)$ results from the polarisation sum 
\be
\sum_{\lambda=1}^2 
e^\mu_{\bk\lambda}e^\nu_{\bk\lambda}=\delta_{\mu\nu}-\frac{k^\mu k^\nu}{k^2}=\delta_{\mu\nu}^{\text tr}(\bk)\;\;.
\la{3.14b}
\ee

There is an important point to deal with before proceeding to the determination of the effective magnetic potential.
The function $\mathcal{C}_{\text{even}}(k,{\tau})$ is continuous  except for the point $\tau=0$ where it has the jump
\be 
\lim_{\tau\to 0}\mathcal{C}_{\text{even}}(k,{\tau})-\mathcal{C}_{\text{even}}(k,0)=\frac{1}{2}\;\;.
\la{7.19a}
\ee
Although this point is of zero measure with respect to the Lebesgue measure, it cannot be disregarded when dealing with stochastic integrals. Indeed, when averaging over loops,
the singular part $\delta(\tau-{\tau}')$ in the covariance 
of stochastic differentials (\ref{3.9a}) will precisely select the value of
$\mathcal{C}_{\text{even}}(k,{\tau}-{\tau}')$ at ${\tau}={\tau}'$. As an illustration, one can consider the $\bX$-average of (\ref{3.14a}) to linear  order in the expansion of the exponential, namely
\begin{align}
&-\beta\sum_{r,s=1}^n\frac{4\pi \hbar e_{\gamma_r}e_{\gamma_s}}{\sqrt{m_{\gamma_r}m_{\gamma_s}}}\int\frac{d^3\bk}{(2\pi)^3}\frac{g^2(k)}{2\omega_\bk}e^{i\bk\cdot(\r_r-\r_s)}\delta_{\mu\nu}^{\text tr}(\bk)\;\;\times\nonumber\\&
\left\langle\int_0^{q_r}dX_r^\mu(\tau)\int_0^{q_s}dX_s^\nu(\tau')
e^{i\bk\cdot(\lambda_{\gamma_r}\bX_r(\tau)-\lambda_{\gamma_s}\bX_s(\tau'))}\right\rangle_\bX\mathcal{C}_{\text{even}}(k,\tilde{\tau}-\tilde{\tau}')\;\;.
\la{3.14}
\end{align}
The average $<\cdots>_\bX$ in (\ref{3.14}) can be calculated by means of the Wick theorem, evaluating all contraction schemes. Contractions involving the product of stochastic differentials yield the term
\begin{align}
&\!\!\!\!\!\int_0^{q_r}\!\!\!\int_0^{q_s}\!\!\!<dX_r^\mu(\tau)dX_s^\nu(\tau')>_\bX
\!\!\left\langle e^{i\bk\cdot(\lambda_{\gamma_r}\bX_r(\tau)-\lambda_{\gamma_s}\bX_s(\tau'))}\right\rangle_\bX\mathcal{C}_{\text{even}}(k,\tilde{\tau}-\tilde{\tau}')\nonumber\\&
=\delta_{rs}\delta_{\mu\nu}\int_0^{q_r}d\tau\int_0^{q_s}d\tau'\(\delta(\tau-\tau')
-\frac{1}{q_r}\)\nonumber\\
&\quad\quad\quad\quad\quad\quad\times\left\langle e^{i\bk\cdot(\lambda_{\gamma_r}\bX_r(\tau)-\lambda_{\gamma_s}\bX_s(\tau'))}\right\rangle_\bX
\mathcal{C}_{\text{even}}(k,\tilde{\tau}-\tilde{\tau}')\;\;.
\la{3.15}
\end{align}
In view of (\ref{7.19a}) the contribution of $\delta(\tau-{\tau}')$ 
in (\ref{3.15}) is
\begin{align}
\delta_{rs}\delta_{\mu\nu} q_r\mathcal{C}_{\text{even}}(k,0)=
\delta_{rs}\delta_{\mu\nu}\lim_{\tau\to\tau'}\mathcal{C}_{\text{even}}(k,\tau-\tau')-\delta_{rs}\delta_{\mu\nu}\frac{q_r}{2}\;\;.
\la{3.16}
\end{align}
Then the contribution of the last term of (\ref{3.16}) to the complete expression (\ref{3.14}) gives
 \begin{align}
 \beta\sum_{r=1}^n q_r\left[\frac{2\pi\hbar }{c}\frac{e^2_{\gamma_{r}}}{m_{\gamma_{r}}}\(\frac{1}{R^3}\sum_{\bk}\frac{g^2(k)}{k}\)\right]=\beta \sum_{r=1}^n
q_rd_{\gamma_{r}}=\beta D_N\;\;.
\la{3.17}
\end{align}
The last line follows from the fact that we have $n$ loops, each of them containing $q_r$ particles of species $\gamma_{r}$, so that 
$D_N$ is the constant (\ref{7.5a}) arising from the normal order rule in the bosonic integral. At linear order, this constant exactly compensates
the term $-\beta D_N$ occuring in the exponent of the total Gibbs weight  
(\ref{3.10}). We conclude from this observation and from (\ref{3.15}) that we can as well use the continuous extension of  $\mathcal{C}_{\text{even}}(k,{\tau})$  (\ref{3.13}) to $\tau=0$
and suppress the constant $D_N$ in (\ref{3.10}), (\ref{3.10b}). A  
proof that this statement holds for all orders is given in Appendix A.

We can now cast the field average (\ref{3.12}) in final form 
\begin{align}
 e^{-\beta D_N}\langle\exp[-\beta U_\A(\LL_1,\ldots,\LL_n)]\rangle_\text{rad}  =&\prod_{r=1}^{n}\exp\(-\frac{\beta
        e_{\gamma_{r}}^{2}}{2}\mathcal{W}_\text{m}(\cl_r,\cl_r)\)\la{7.18a}\\
    &\times\exp\(-\beta\sum_{r<s}^{n}e_{\gamma_{r}}e_{\gamma_{s}}\mathcal{W}_\text{m}(\cl_r,\cl_s)\)\;\;.
\nonumber    
\end{align}
Here we have introduced the effective magnetic potential
\begin{align}
  &  \mathcal{W}_\text{m}(\cl_r,\cl_s)=\frac{1}{\beta
        \sqrt{m_{\gamma_r}m_{\gamma_s}}c^{2}}\int\!\!\! \frac{\d\bk }{(2\pi)^{3}}\,
    \e^{i\bk\cdot(\r_r-\r_s)}\la{7.18}\\\times
    &\int_{0}^{q_{r}}\!\!\! \d X_{r}^\mu(\tau)\,
    \e^{-i\bk\cdot\lambda_{\gamma_{r}} \bX_{r}(\tau)}\! \int_{0}^{q_{s}}\!\!\! \d
    X_{s}^\nu(\tau')\,
    \e^{i\bk\cdot\lambda_{\gamma_{s}}\bX_{s}(\tau')}\
   \frac{4\pi g^2(k)}{k^2}\delta_{\mu\nu}^{\text tr}(\bk){\cal Q}(k,\tilde{\tau}-\tilde{\tau}')\;. \nonumber
\end{align}
To obtain (\ref{7.18a}) and (\ref{7.18}), we have separated in (\ref{3.14a}) the terms
$r=s$ refering to the self energies of loops from the terms $r\neq s$ giving rise to pairwise loop interactions. The function 
\begin{align}
{\cal Q}(k,{\tau})&=\frac{\lambda_{\text{ph}}k}{2\sinh( \lambda_{\text{ph}}k/2)}\cosh[\lambda_{\text{ph}}k(|\tau|-1/2)]\nonumber
 \\
&=\(\frac{\lambda_{\text{ph}}k}{2}\)\frac{e^{\lambda_{\text{ph}}k(|\tau |-1)}+e^{-\lambda_{\text{ph}}k|\tau|}}{1-e^{-\lambda_{\text{ph}}k}}, \quad |\tau| \leq 1
\la{7.19}
\end{align}
is, up to the factor $\lambda_{\text{ph}}k$, the even part (\ref{3.13}) of the covariance of the free photon field written in terms of the photon thermal wave length $\lambda_{\text{ph}}=\beta\hbar c$. In view of the discussion following (\ref{7.19a}) and the result of Appendix A, it is understood that this function is given by the formula (\ref{7.19}) including the point $\tau=0$   and the factor $e^{-\beta D_N}$ has been cancelled in the right hand side
of (\ref{7.18a}). The $\tau$-periodic function ${\cal Q}(k,\tilde{\tau})$, ${\cal Q}(k,0)={\cal Q}(k,1)$, is normalized in such a way that it equals one when the electromagnetic field is classical 
\be
\lim_{\lambda_{\text{ph}}\to 0}{\cal Q}(k,{\tau})=1\;\;.
\la{7.20}
\ee
In this limit, the magnetic potential $ \mathcal{W}_\text{m}(\cl_r,\cl_s)$ reduces to formula (82) of I where radiation has been treated classically. Hence all effects due to the quantum nature of the photon field are contained in the sole function ${\cal Q}(k,{\tau})$.

The Gaussian integration of the radiation field has provided the sum of pair potentials \eqref{7.18a} between loops as in Paper I. Then, thermal averages of particle observables calculated with the normalized reduced density matrix $\rho_{L,R}$ \eqref{B.4a} have a simple structure when expressed in the system of loops. Combining \eqref{3.10b} and \eqref{7.18a}, one forms the complete effective Gibbs weight (up to normalisation)
\begin{align}
&\exp\Big[-\beta\(\sum_{r=1}^n U(\LL_r) + \frac{e_{\gamma_r}^2}{2} \mathcal{W}_\text{m}(\LL_r,\LL_r)\)\Big]\nonumber
\\ &\times\exp\Big[-\beta \Big(U_\text{pot}(\LL_1,\ldots,\LL_n) + \sum_{r<s}^{n}e_{\gamma_{r}}e_{\gamma_{s}}\mathcal{W}_m(\cl_r,\cl_s) \Big)\Big] \la{7.21}
\end{align}
comprising one-loop and two-loop interactions. This structure allows the use of standard diagrammatic methods of classical statistical mechanics, like Mayer graph expansions. This is illustrated in the next section, where large-distance asymptotic particle correlations are investigated. 

Note that as in Paper I, it is unlikely that $\rho_{L,R}$ can be cast in a convenient operator form $\rho_{L,R} \propto \e^{-\beta H_\text{eff}(\{\mathbf{p}_i, \r_i\})}$ depending on the original quantum-mechanical momenta and positions $\{\mathbf{p}_i, \r_i\}$ of the particles. Again, the magnetic interaction $\mathcal{W}_\text{m}$ \eqref{7.18} is a two-times functional of the Brownian loops reflecting the photonic bath environment. It lacks the equal-time constraint necessary to come back to a simple operator form by using the Feynman--Kac--It\^o formula backwards \cite{Feynman-Hibbs}.

\section{Asymptotic particle correlations} 

We determine the behaviour of the particle density correlation in presence of the thermalized quantum electromagnetic field in the two regimes (\ref{8.1a}) and (\ref{8.1b}) discussed in the introduction. 

\subsection {\!\!\!Partial\! screening\! of the\! Coulomb\! interaction \\ by thermal photons in the range $\lambda_\text{mat}\! \ll \!\lambda_\text{ph} \!\ll \!\!r$} 
In the regime \eqref{8.1a}, $r$ is larger than any typical length of the model. The asymptotic analysis of the correlation is based on the large-distance behaviour of the part of the interaction
\footnote{Exchange effects are short ranged and play no role here. Only one particle loops, i.e. filaments, are considered. }
\begin{align} 
    \mathcal{W}({\cal F}_a,{\cal F}_b)= 
    \mathcal{W}_c({\cal F}_a,{\cal F}_b) + 
    \mathcal{W}_\text{m}({\cal F}_a,{\cal F}_b), 
    \la{8.2} 
\end{align} 
which is responsible for the power-law decay. In this formula, $\mathcal{W}_c({\cal F}_a,{\cal F}_b)$ is the residual interaction (due to quantum fluctuations) that is left when Coulomb divergencies are resummed in Mayer graphs (see formula (28) of I). It has the asymptotic dipolar form 
\begin{align} 
    &\mathcal{W}_c({\cal F}_a,{\cal F}_b)\sim\quad\quad\quad\quad\quad\quad |\r_{a}-\r_{b}|\to\infty \nonumber\\& 
     \int_{0}^{1}\!\!\! \d s_{a}\!\! \int_{0}^{1}\!\!\! \d s_{b}\, 
    (\delta(s_{a}\!-\!s_{b})\!-\!1)\(\lambda_{\gamma_{a}}\b\xi_{a}(s_{a})\cdot 
    \nabla_{\r_{a}} \)\( 
    \lambda_{\gamma_{b}}\b\xi_{b}(s_{b})\cdot\nabla_{\r_{b}}\) 
    \frac{1}{|\r_{a}-\r_{b}|}\;. 
 \la{8.3} 
\end{align} 
It turns out that the large-distance asymptotics of $\mathcal{W}_\text{m}({\cal F}_a,{\cal F}_b)$, determined by the small-$k$ behaviour of the integrand of (\ref{7.18}), are also dipolar. Indeed, we first observe that ${\cal Q}(k,{\tau})$ is an analytic function of $\bk$ and has the small-$k$ expansion
\be 
{\cal Q}(k,{\tau})=1 +\frac{(\lambda_{{\rm ph}}k)^2}{2} \left[\tau^2-|\tau| +\frac{1}{6}\right] +{\cal O}((\lambda_{{\rm ph}}k)^4)\;. 
 \la{8.4} 
\ee 
Inserting this in (\ref{7.18}) gives
\begin{align} 
  \mathcal{W}_\text{m}&({\cal F}_a,{\cal F}_b) \sim W_\text{m}({\cal F}_a,{\cal F}_b) 
  - \frac{2\pi\lambda_{{\rm ph}}^2}{\beta 
        \sqrt{m_{a}m_{b}}c^{2}}\;\int\!\!\! \frac{\d\bk }{(2\pi)^{3}}\, 
    \e^{i\bk\cdot(\r_a-\r_b)}\frac{ k^\mu k^\nu}{k^2}\nonumber\\ 
    &\times\int_{0}^{1}\!\!\! \d \xi_{a}^\mu(\tau)\, 
  \! \int_{0}^{1}\!\!\! \d 
   \xi_{b}^\nu(\tau')\,\left[(\tau-\tau')^2-|\tau-\tau'| +\frac{1}{6}\right] 
 ,\quad |\r_{a}-\r_{b}|\to\infty\;. 
\la{8.5} 
\end{align} 
The first term in the r.h.s of (\ref{8.4}) leads back to the effective magnetic potential $W_\text{m}$ associated to the classical electromagnetic field (formula (22) of I). In the second term, the $k^{-2}$ factor in the integrand of (\ref{7.18}) has been cancelled by the term of second order in $k$ of (\ref{8.4}) and we have set $\bk=\mathbf{0}$ in the exponentials of the paths $\bxi_a(\tau)$ and $\bxi_b(\tau')$. In this way, we have retained the lowest order singular part in $\bk$ in the last term of (\ref{8.5}). This part is $-k^\mu k^\nu/k^2$ coming from the transverse delta function  (\ref{3.14b}). The double stochastic integral in (\ref{8.5}) is calculated with the result 
\begin{align} 
&\int_0^1\!\!\!\d\tau\!\!\int_0^1\!\!\!\d\tau'\ \xi_a^\mu(\tau)\xi_b^\nu(\tau')\  
\frac{\partial^2}{\partial\tau\partial\tau'}\left[(\tau-\tau')^2-|\tau-\tau'| +\frac{1}{6}\right]\nonumber\\& 
=2\int_0^1\!\!\!\d\tau\!\!\int_0^1\!\!\!\d\tau'\,(\delta(\tau-\tau')-1)\ \xi_a^\mu(\tau) 
\xi_b^\nu(\tau')\;. 
\la{8.6} 
\end{align} 
In virtue of the identity 
\be 
\frac{\lambda_{{\rm ph}}^2}{\beta 
        \sqrt{m_{a}m_{b}}c^{2}}=\lambda_a\lambda_b 
\la{8.7} 
\ee 
equation (\ref{8.5}) eventually reads 
\begin{align}
&\mathcal{W}_\text{m}({\cal F}_a,{\cal F}_b) \sim W_\text{m}({\cal F}_a,{\cal F}_b) \la{8.8}
\\ &-\lambda_a\lambda_b\int_0^1\!\!\!\d\tau\!\!\int_0^1\!\!\!\d\tau'\,(\delta(\tau-\tau')-1)\
\xi_a^\mu(\tau) \xi_b^\nu(\tau') 
\int\!\!\! \frac{\d\bk }{(2\pi)^{3}}\, 
    \e^{i\bk\cdot(\r_a-\r_b)}\frac{4\pi k^\mu k^\nu}{k^2} \;.\nonumber
\end{align}
Performing the Fourier transform, we see that, up to the sign, the second term in the r.h.s. of \eqref{8.8} is identical to the asymptotic tail (\ref{8.3}) of $\mathcal{W}_c$. The latter is therefore exactly cancelled in the total interaction $ \mathcal{W}({\cal F}_a,{\cal F}_b)= \mathcal{W}_c({\cal F}_a,{\cal F}_b) + \mathcal{W}_\text{m}({\cal F}_a,{\cal F}_b) $ as $|\r_a-\r_b|\to\infty$. We conclude that in the region $r\gg \lambda_{{\rm ph}}$  the dominant part of this algebraic Coulombic tail is screened by thermalized photons. The  tail of the interaction 
\be 
\mathcal{W}({\cal F}_a,{\cal F}_b)\sim W_\text{m}({\cal F}_a,{\cal F}_b), 
\quad  |\r_a-\r_b|\to\infty 
\la{8.9} 
\ee 
reduces therefore to the pure unscreened effective magnetic current-current interaction $W_\text{m}({\cal F}_a,{\cal F}_b)$ induced by the classical field, whose asymptotic dipolar form is given by formula (25) of I. 

We can now follow the asymptotic analysis presented in Section V of Paper I to show that the tail of the correlation exhibits again a generic $r^{-6}$ decay. All statements made there regarding the magnetic potential with the classical field $W_\text{m}$ hold for the magnetic potential with the quantum field $\mathcal{W}_\text{m}$. The transversality argument used to show the vanishing of the convolution element (I.50) works identically provided that the rotationally invariant function $\mathcal{Q}(k,s_2-s_b)$ \eqref{7.19} is included in the definition of the tensor $T^{\nu_2}(\k,s_1,s_b)$ (I.51). This tensor still transforms in a covariant manner under rotations of $\k$, so that its contraction with the transverse delta function cancels (I.50). Similar modifications done in the other convolution elements mentionned after Eq. (I.51) imply that $\mathcal{W}_\text{m}$ does not contribute to the $\mathcal{W}$-convolution chains occurring in (I.49). The dipolar character of the large-distance interaction $\mathcal{W}$ then ensures that the correlation function decays as $r^{-6}$. However, the amplitude of this decay is now affected by the partial screening \eqref{8.9} which is due to the quantum nature of the photonic bath. 

In order to illustrate this point, let us determine the coefficient of the $r^{-6}$ decay at lowest order in $\hbar$. Proceeding word for word as in Section V of Paper I, one sees that this decay is eventually governed by 
\begin{align} 
    \tfrac{1}{2}\big[-\beta e_{\gamma_1} e_{\gamma_2}\mathcal{W} (\F_1,\F_2)\big]^2 \la{8.9a}
\end{align}
with root points dressed by classical correlations and evaluated at lowest order in $\hbar$. Since $\mathcal{W}=\mathcal{W}_c+\mathcal{W}_\text{m}$ depends on $\hbar$ solely through the couplings $\lambda_\text{mat} k$ in $\mathcal{W}_c$, and $\lambda_\text{mat} k$, $\lambda_\text{ph} k$ in $\mathcal{W}_\text{m}$, evaluating these potentials at lowest order in $\hbar$ amounts exactly to selecting their large-distance ($k\to 0$) asymptotic behaviour. The Coulombic dipolar tail of $\mathcal{W}_c$ is therefore cancelled by the photon-induced partial screening \eqref{8.9}, and the large distance behaviour of the two-particle truncated correlation in the semi-classical regime (high-temperature or lowest order in $\hbar$) reads:
\begin{align} 
        \rho_\text{T}(\gamma_a,\r_a,\gamma_b,\r_b) \sim &  \frac{\hbar^4 
        \beta^4}{48 }\sum_{\gamma_1,\gamma_2} \left[ \int\!\! \d 
    \r\ n_\text{T}^\text{cl}(\gamma_a,\gamma_1,\r)\right] \left[ \int\!\! \d \r 
    \ n_\text{T}^\text{cl}(\gamma_2,\gamma_b,\r)\right]\nonumber\\ 
        &\times \frac{e_{\gamma_1}^2 e_{\gamma_2}^2}{\beta m_{\gamma_1}c^2 \beta m_{\gamma_2}c^2}\;\; 
   \frac{1}{|\r_a-\r_b|^{6}}.
\label{8.10} 
\end{align} 
This corresponds to omit the Coulombic part of the correlation calculated in I, Formula (53). Only the current-current interaction induced by the thermal motion of the particles contributes to the tail \eqref{8.10} in the regime $r \gg \lambda_\text{ph}$.

The analysis of the particle-charge and charge-charge correlation function can be performed in the same way. As recalled in (I.46) and (I.47), the Mayer bonds are built from a rapidly decaying resummed potential $\Phi_{{\rm elec}}$ and the quantum asymptotically dipolar potential $\mathcal{W}$ (\ref{8.2}).
When the charge observable is considered, following the dressing method described in Sect. VI.A.3 of \cite{BrMa}, one sees that an additional screening factor involving $\Phi_{{\rm elec}}$ occurs at the root points of the graphs. This generically weakens the decay of the particle-charge correlation to $r^{-8}$ and that of the charge-charge correlation to  $r^{-10}$. As for the particle-particle correlation (\ref{8.10}), the amplitudes of the tails are again determined by $\mathcal{W}^2$. Because of the asymptotic cancellation of the Coulombic part in $\mathcal{W}$ (see (\ref{8.9})), these amplitudes also inherit the small relativistic factor $(\beta m c^2)^{-2}$.

\subsection{Predominance of electrostatic correlations in the range $\lambda_\text{mat} \ll r \ll \lambda_\text{ph}$} 

Let us now focus on the second regime, $\lambda_{{\rm mat}}\ll r\ll \lambda_{{\rm ph}}$, i.e. we consider the correlation between particles that are separated by distances much smaller than the wavelength of thermalized photons. We first give a rough estimate of the order of magnitude of $ \mathcal{W}_\text{m}({\cal F}_a,{\cal F}_b)$
relative to $ \mathcal{W}_c({\cal F}_a,{\cal F}_b)$. In this aim it is convenient to scale the Fourier variable as $\bk\to\bk/r,\;\r=\r_a-\r_b,$ yielding in (\ref{7.18})
\begin{align}
  &  \mathcal{W}_\text{m}({\cal F}_a,{\cal F}_b)=\frac{1}{\beta
        \sqrt{m_{a}m_{b}}c^{2}}\frac{1}{r}\int\!\!\! \frac{\d\bk }{(2\pi)^{3}}\,
    \e^{i\bk\cdot\hat{\r}}    \int_{0}^{1}\!\!\! \d \xi_{a}^\mu(\tau)\,
    \e^{i\bk\cdot(\lambda_{a}/r)\b\xi_{a}(\tau)}\!\nonumber\\ &\times\int_{0}^{1}\!\!\! \d
    \xi_{b}^\nu(\tau')\,
    \e^{-i\bk\cdot(\lambda_{b}/r)\b\xi_{b}(\tau')}\
   \frac{4\pi g^2(k/r)}{k^2}\delta_{\mu\nu}^{\text tr}(\bk){\cal Q}(k/r,\tau-\tau'),\quad     \hat{\r}=\frac{\r}{r}\;\;. 
 \la{8.11}
\end{align}
Since $ \lambda_{{\rm ph}}/r$ is now a large number, it is not allowed to expand \\${\cal Q}(k/r,\tau-\tau')$ for small $\bk$, but we note from (\ref{7.19}) that this function is of the form
$\lambda_{{\rm ph}}/r$ times a bounded function of $\lambda_{{\rm ph}}/r$. Therefore ${\cal Q}(k/r,\tau-\tau')$ cannot grow faster than $\lambda_{{\rm ph}}/r$. Then the order of magnitude   
of  $\mathcal{W}_\text{m}$ is at most
\be
 \mathcal{W}_\text{m}=\frac{1}{\beta\sqrt{m_{a}m_{b}}c^{2}r}{\cal O}\(\frac{\lambda_{{\rm ph}}}{r}\)\;\;.
\la{8.12}
\ee
On the other hand, one sees from (\ref{8.3}) that the order of magnitude of $ \mathcal{W}_c$ for $r\gg\lambda_{{\rm mat}}$
is
\be
\mathcal{W}_c\sim\frac{\lambda_a\lambda_b}{r^3}\;.
\la{8.13}
\ee
Combining (\ref{8.12}) and (\ref{8.13}) together with (\ref{8.7})  gives
\be
 \mathcal{W}_\text{m}=\mathcal{W}_c\;\;{\cal O}\(\frac{r}{\lambda_{{\rm ph}}}\)
\la{8.14}
\ee
Hence, in the range (\ref{8.1b}), the total
interaction 
\be
\mathcal{W}=\mathcal{W}_c+ \mathcal{W}_\text{m}=\mathcal{W}_c\(1+{\cal O}\(\frac{r}{\lambda_{{\rm ph}}}\)\)
\la{8.14a}
\ee
is given by its Coulombic part up to a small correction. It is therefore expected that all predictions on correlation decays
are the same as those derived from pure electrostatics up  to  terms that vanish as $r/ \lambda_{{\rm ph}}\to 0$. 
This reasoning is mathematically not complete since when (\ref{8.11}) is used as a bond in Mayer graphs,  
loop averages and wave number Fourier 
integrals have to be performed first and shown to yield finite values. As an example we establish in  Appendix B the precise estimate
\be
\left\langle \mathcal{W}_\text{m}^2(\F_a,\F_b)\right\rangle
_{\b\xi_a,\b\xi_b}\sim 120A\left\langle \mathcal{W}_c^2(\F_a,\F_b)\right\rangle_{\b\xi_a,\b\xi_b}\(\frac{r}{\lambda_{{\rm ph}}}\)^3,\quad A<\infty
 \la{8.16}
\ee
as $\lambda_{\rm cut}/r\to 0,\;\lambda_a/r,\lambda_b/r\to 0,\;\lambda_{{\rm ph}}/r\to\infty$, implying 
\be
\left\langle \mathcal{W}_\text{m}^2(\F_a,\F_b)\right\rangle
_{\b\xi_a,\b\xi_b}=\left\langle \mathcal{W}_c^2(\F_a,\F_b)\right\rangle_{\b\xi_a,\b\xi_b}{\cal O}\(\(\frac{r}{\lambda_{{\rm ph}}}\)^3\)
 \la{8.17}
\ee
in the range (\ref{8.1b}).
Thus the square fluctuation of  $\mathcal{W}_\text{m}$ (entering e.g. in (\ref{8.9a}) for the evaluation of the correlation) is negligible compared to that of $ \mathcal{W}_c$.

\section{Concluding remarks}

In this paper, we have presented a formalism adapted to the study of non relativistic matter in thermal equilibrium with the photon field. In the joint functional integral representation of matter and field, the field variables can be integrated out, yielding an effective classical-like statistical description of the state of matter.
As a first application, we have shown that the cloud of thermalized photons participates in the screening of the Coulomb potential
by supressing the dipolar electric contribution to the $r^{-6}$ tail of the particle correlations, as illustrated in \eqref{8.10}. In electrolytes at room temperature, both the de Broglie and classical Debye lengths are in the range of $10^{-10}\text{m}$ (a few Angstr\"oms). Moreover, the parameter $\sqrt{\beta \bar{m} c^2}$ is of order $\approx 10^5$ so that $\lambda_\text{ph} \approx 10^{-5} \text{m}$, see (\ref{8.1}). It is known from \cite{alastuey-martin} that the crossover between Debye-H\"uckel (exponential) screening and quantum (algebraic) screening occurs at distances of about $60$ times the Debye screening length. Consequently, the further reduction of the amplitude of the correlation tail by photon screening occurs at even much larger distances and with an exceedingly small amplitude. This makes the phenomenon probably hardly observable in such systems.
At the conceptual level, it is however an interesting effect of the thermal radiation that, to our knowledge, has not been exhibited in the literature before.

The effective magnetic potential $\mathcal{W}_\text{m}$ (\ref{7.18}) defined in Section 5 embodies in an exact manner orbital diamagnetic interactions, namely interactions between currents due to thermal motion of charges. This current-curent interaction is at the origin of the correlation tail (\ref{8.10}). 
The order of magnitude of $\mathcal{W}_\text{m}$ is by a factor $(\beta \bar{m}c^2)^{-1}$ smaller than that of the electrostatic potential. One should  however be aware that $\mathcal{W}_\text{m}$ is not the unique source of relativistic effects. A preliminary investigation  \cite{Samidiplome} shows that the Pauli coupling terms of spins with the field contribute to the correlation tail at the same order $(\beta \bar{m}c^2)^{-2}$ as that found in (\ref{8.10}) as a consequence of pure orbital magnetism. Moreover, the non relativistic form of the particles kinetic energy in the Hamiltonian   
(\ref{B.1}) has itself $c^{-2}$ corrections (e.g., spin-orbit interaction, Darwin term) that will likely contribute to the asymptotic form of the particle correlations. Hence a complete determination of the particle correlations tails at order $(\beta \bar{m}c^2)^{-2}$ will require further investigations.  

The tools developed in this paper lend themselves to a detailed microscopic study of various problems.
In view of the brief report \cite{Janco} questioning the findings of paper I on electromagnetic fluctuations, we aim to revisit the problem in the case of a quantized field. 
Thermal broadening of spectral lines and retardation effects on van der Waals forces between recombined atoms or molecules in a medium at finite density and temperature could conveniently be studied within this formalism.
Indeed, the latter situations involve quantum mechanical binding which is not perturbative in the matter-field coupling constant.  
Standard many-body Feynman diagram techniques would necessitate infinite resummations to describe  bound state formation, whereas cluster expansions in the form presented in \cite{ScreenedCluster} (properly generalized to include the full electromagnetic coupling) give a direct access to recombined entities together with their interaction with the radiation field. Finally the theory of the Casimir effect has received much attention recently. It is now conceivable to elaborate a full microscopic theory of this effect by extending the analysis presented in \cite{BuMa} to TQED. We plan to address these questions in future works.

\vspace{4mm}

\noindent{\Large Appendix A}

\vspace{4mm}
As seen in the first order calculation leading to (\ref{3.17}), the compensation of the constant $D_N$ comes from the particle self energies $r=s$. It is therefore appropriate to  
single out in the exponent of (\ref{3.14a}) a diagonal $r=s$ term and write its $\bX_r$ average as (dropping now the particle index $r$)
\begin{align}
&I =\left\langle \exp\left[-\beta\int\frac{d^3\bk}{(2\pi)^3}
\int_0^{q}dX^\mu(\tau)\int_0^{q}dX^\nu(\tau')\right.\right.
\nonumber\\&\times
\left.\left.\phantom{\int}e^{-i\bk\cdot\lambda(\bX(\tau)-\bX(\tau')}
{\Gamma}_{\mu\nu}(k,\tilde{\tau}-\tilde{\tau}')\right]F(\bX)\right\rangle_\bX\;,
\la{A1}
\end{align}
where we have set for brevity
\be
{\Gamma}_{\mu\nu}(k, {\tau})=\frac{4\pi\hbar e^2}{m} \frac{g^2(k)}{2\omega_\bk}\delta_{\mu\nu}^{\text tr}(\bk)\mathcal{C}_{\rm{even}}(k,{\tau})\;\;.
\la{A2}
\ee
In (\ref{A1}) $F(\bX)$ is a functional of $\bX$ containing all possible other dependences of $\bX$ in (\ref{3.14a}). 
Expanding the exponential in (\ref{A1}) gives
\begin{align}
I&=\sum_{n=1}^\infty \frac{(-\beta)^n}{n\;!}
\(\prod_{j=1}^n\int\frac{d^3\bk_j}{(2\pi)^3}\)\nonumber
\\&\times
\left\langle\(\prod_{j=1}^n\int_0^{q}dX^{\mu_j}(\tau_j)\int_0^{q}
dX^{\nu_j}(\tau'_j)e^{-i\bk_j\cdot\lambda(\bX(\tau_j)-\bX(\tau'_j)}\)F(\bX)\right\rangle_\bX\nonumber\\
&\quad\quad\quad\quad\quad\quad\quad\quad\quad\quad\quad\quad\times\(\prod_{j=1}^n{\Gamma}_{\mu_j\nu_j}(k_j,\tilde{\tau_j}-\tilde{\tau_j})\)\;\;.
\la{A3}
\end{align}
We call a matched contraction the contraction of a pair of stochastic differentials $<dX^{\mu_j}(\tau_j)dX^{\nu_j}(\tau'_j)>_\bX=\delta_{\mu_j\nu_j}\(\delta(\tau_j-\tau'_j)-1/q\)d\tau_jd\tau'_j$,\\ where times have the same index $j$. It is clear that the $\delta(\tau_j-\tau'_j)$ occuring in matched contractions will evaluate ${\Gamma}_{\mu_j\nu_j}(k_j,\tilde{\tau_j}-\tilde{\tau_j})$ at $\tau_j=\tau'_j$. Such matched contractions can only arise from the product $\prod_ {j=1}^ndX^{\mu_j}(\tau_j)  dX^{\nu_j}(\tau'_j)$  in (\ref{A3}).   Contraction between a differential from this product with a differential occuring in $F(\bX)$, or contractions within $F(\bX)$, will always involve two time arguments belonging to different $\mathcal{C}_{\rm{even}}$ functions. They are of the type
\begin{align}
&\left\langle\int_0^q dX^\mu(\tau)\int_0^q dX^\nu(\sigma)
{\Gamma}_{\mu\mu'}(k,\tilde{\tau}-\tilde{\tau}')
{\Gamma}_{\nu\nu'}(k',\tilde{\sigma}-\tilde{\sigma}')\right\rangle_\bX
=\nonumber\\&
\int_0^q d\tau\int_0^q d\sigma\(\delta(\tau-\sigma)-\frac{1}{q}\)
{\Gamma}_{\mu\mu'}(k,\tilde{\tau}-\tilde{\tau}')
{\Gamma}_{\mu\nu'}(k',\tilde{\sigma}-\tilde{\sigma}')=
\nonumber\\&
q\!\left[\int_0^1d\tau{\Gamma}_{\mu\mu'}(k,{\tau}\!-\!\tilde{\tau}')
{\Gamma}_{\mu\nu'}(k',{\tau}\!-\!\tilde{\sigma}')\!-\!\int_0^1\!\!d\tau
{\Gamma}_{\mu\mu'}(k,{\tau})\!\!\int_0^1\!\!d\sigma{\Gamma}_{\mu\nu'}(k',{\sigma})\right]\;\;.
\la{A3a}
\end{align}
For such contractions, $\mathcal{C}_{\rm{even}}(k,{\tau})$ can be treated as a continuous function everywhere since in integrals of the type (\ref{A3a}) the discontinuity
(\ref{7.19a}) at the single point $\tau=0$ is irrelevant. 

To evaluate the $\bX$ average in (\ref{A3}), we select therefore terms having exactly $m$ matched contractions, $0\leq m\leq n$. Because of the invariance of the product under exchange of its factors there are \\$n\;!/m\; !(n-m)\; !$ such terms giving the same contribution. This leads to
\begin{align}
I&=\sum_{n=1}^\infty \frac{(-\beta)^n}{n\;!}
\sum_{m=0}^n\frac{n\; !}{m\; !(n-m)\; !}\nonumber\\&\times
\prod_ {j=1}^m\int\frac{d\bk_j}{(2\pi)^3}\left[\int_0^qd\tau_j \int_0^q d\tau'_j
{\Gamma}_{\mu\mu}(k_j,\tilde{\tau_j}-\tilde{\tau_j}')\(\delta(\tau_j-\tau_j')-\frac{1}{q}\)\right]\nonumber\\&
\quad\quad\quad\quad\times\left\langle\(\prod_ {j=1}^m e^{-i\bk_j\cdot\lambda(\bX(\tau_j)-\bX(\tau'_j))}\)(B(\bX))^{n-m}F(\bX)\right\rangle_{{\rm unmatched}}
\la{A4}
\end{align}
with
\begin{align}
&B(\bX)=\nonumber\\ &\int\frac{d\bk}{(2\pi)^3}\int_0^{q}dX^\mu(\tau)\int_0^{q}dX^\nu(\tau')e^{-i\bk\cdot\lambda\bX(\tau)-\bX(\tau'))}{\Gamma}_{\mu\nu}(k,\tilde{\tau}-\tilde{\tau}')\;\;.
\la{A5}
\end{align}
The square bracket in (\ref{A4}) is the result of $m$ matched contractions. In the average $<\cdots>_{{\rm unmatched}}
$, all matched contractions are omitted.
In (\ref{A4}), we further expand the product of $\(\delta(\tau_j-\tau_j')-1/q\)$ factors and perform the $\delta$ function integrations leading to
 \begin{align}
I&=\sum_{n=1}^\infty \frac{(-\beta)^n}{n\;!}
\sum_{m=0}^n\frac{n\; !}{m\; !(m-n)\; !}\sum_{l=0}^m\frac{m\; !}{l\; !(m-l)\; !}\nonumber\\&\times
\(q\int\frac{d\bk}{(2\pi)^3}{\Gamma}_{\mu\mu}(k,0)\)^l\left\langle (B(\bX))^{n-m}(D(\bX))^{m-l}F(\bX)\right\rangle_{{\rm unmatched}}
\la{A7}
\end{align}
with
 \begin{align}
D(\bX)\!=\!\!-\!\frac{1}{q}\!\int\frac{d\bk}{(2\pi)^3}\int_0^{q}d\tau\int_0^{q}d\tau'e^{-i\bk\cdot\lambda(\bX(\tau)-\bX(\tau'))}\;
{\Gamma}_{\mu\mu}(k,\tilde{\tau}\!-\!\tilde{\tau}')\;.
\la{A8}
\end{align}
Finally, rearranging the sums yields
 \begin{align}
I=\exp\(-\beta q\int\frac{d\bk}{(2\pi)^3}{\Gamma}_{\mu\mu}(k,0)\)
\left\langle e^{-B(\bX)}e^{-D(\bX)}F(\bX)\right\rangle_{{\rm unmatched}}\;\;.
\la{A9}
\end{align}
It is seen from the definition (\ref{A2}) that $\beta q\int\tfrac{d\bk}{(2\pi)^3}{\Gamma}_{\mu\mu}(k,0)$ is equal to the constant $-qd$ 
plus the contribution of $\mathcal{C}_{\rm{even}}(k,{\tau})$ extended by continuity at $\tau=0$, exactly as in (\ref{3.15})-(\ref{3.17}). Since the loop shapes $\bX_r, \; r=1,\ldots, n$, are independent random variables, the same calculation can  successively be carried out
for $n$ loops, providing
a factor $e^{\beta D_N}$ that cancels the factor $e^{-\beta D_N}$ due to normal ordering in (\ref{3.10b}).  Performing the procedure
(\ref{A1})-(\ref{A9}) backwards after this cancellation thus shows the validity of formula (\ref{7.18a}), where the effective magnetic potential $\mathcal{W}_{\text m}$ (\ref{7.18}) is defined with the continuous function  $\mathcal{Q}(k,\tau)$ (\ref{7.19}) for all $\tau$. 
\vspace{4mm}

\noindent {\Large Appendix B}

\vspace{4mm}

 The $\bsl{\xi}_a,\bsl{\xi}_b$  average of $\mathcal{W}_{\text m}^{2}$ reads in terms of the scaled variables $\bsl{q}_{a}=\bsl{k}_{a} r$ and $\bsl{q}_{b}=\bsl{k}_{b} r$ 
\begin{align}\la{C.1}
&\<\mathcal{W}_{\text m}^{2}({\cal F}_a,{\cal F}_b)\>_{\bsl{\xi}_a,\bsl{\xi}_b}=
\frac{1}{(\beta m_{a} c^{2})(\beta m_{b} c^{2})}\frac{1}{r^2}F\left(\frac{\lambda_{\rm cut}}{r},\frac{\lambda_{a}}{r}, \frac{\lambda_{b}}{r},\frac{\lambda_{\rm ph}}{r}\right)\;,
\end{align}
where we have introduced the function of dimensionless parameters
\begin{align}
&F\left(\frac{\lambda_{\rm cut}}{r},\frac{\lambda_{a}}{r},\frac{\lambda_{b}}{r},\frac{\lambda_{\rm ph}}{r}\right)=
\int_{q_1\leq \frac{r}{\lambda_{\rm ph}}} \frac{d\bsl{q}_{1}}{(2\pi)^3}\int_{q_2\leq \frac{r}{\lambda_{\rm ph}}}\frac{d\bsl{q}_{2}}{(2\pi)^3}
e^{i(\bsl{q}_{1}+\bsl{q}_{2})\cdot\hat{\bsl{r}}}
\nonumber\\ &\times\frac{(4\pi)^2}{q_{1}^{2}q_{2}^{2}}\delta_{\mu\nu}^{tr}(\bsl{q}_{1})\delta_{\epsilon\delta}^{tr}(\bsl{q}_{2})
\left\langle e^{i\bsl{q}_{1}\cdot(\frac{\lambda_a}{r}\bsl{\xi}_a(\tau)-\frac{\lambda_b}{r}\bsl{\xi}_b(\tau'))}\,e^{i\bsl{q}_{2}\cdot(\frac{\lambda_a}{r}\bsl{\xi}_a(\sigma)-\frac{\lambda_b}{r}\bsl{\xi}_b(\sigma'))}\right.\nonumber\\ &\times
\left.
\!\!\int_0^1\!\!d\xi_{a}^{\mu}(\tau)\!\!\int_0^1\!\!d\xi_{b}^{\nu}(\tau')\!\!\int_0^1\!\!d\xi_{a}^{\epsilon}(\sigma)\!\!\int_0^1\!d\xi_{b}^{\delta}(\sigma'))\right\rangle_{\bsl{\xi}_a,\bsl{\xi}_b}\!\!
\mathcal{Q} \left(\frac{q_1}{r},\tau-\tau'\right)\mathcal{Q}\left(\frac{q_2}{r},\sigma-\sigma'\right).
\la{C.2}
\end{align}
The ultra-violet cut-off functions $g(q_{1})$ and $g(q_{2})$ have been replaced by the appropriate restrictions of the domains of integration. Then
\begin{align}
&F\left(\frac{\lambda_{\rm ph}}{r}\right)=\lim_{\frac{\lambda_{\rm cut}}{r},\frac{\lambda_{a}}{r},\frac{\lambda_{b}}{r}\to 0}F\left(\frac{\lambda_{\rm cut}}{r},\frac{\lambda_{a}}{r},\frac{\lambda_{b}}{r},\frac{\lambda_{\rm ph}}{r}\right)=
\nonumber\\&\int \frac{d\bsl{q}_{1}}{(2\pi)^3}\int\frac{d\bsl{q}_{2}}{(2\pi)^3}\,\,
e^{i(\bsl{q}_{1}+\bsl{q}_{2})\cdot\hat{\bsl{r}}}\,\,\frac{(4\pi)^2}{q_{1}^{2}q_{2}^{2}}\delta_{\mu\nu}^{tr}(\bsl{q}_{1})\delta_{\epsilon\delta}^{tr}(\bsl{q}_{2})\;\;\times
\nonumber\\
&\left[\<\!\!\int_0^1\!\!d\xi_{a}^{\mu}(\tau)\!\!\int_0^1\!\!d\xi_{b}^{\nu}(\tau')\!\!\int_0^1\!\!d\xi_{a}^{\epsilon}(\sigma)\!\!\int_0^1\!\!d\xi_{b}^{\delta}(\sigma'))\>_{\bsl{\xi}_a,\bsl{\xi}_b}\!\!\!
\mathcal{Q} \!\left(\frac{q_1}{r},\tau\!-\!\tau'\right)\mathcal{Q}\!\left(\frac{q_2}{r},\sigma\!-\!\sigma'\right)\!\right].
\la{C.3}
\end{align}
The $\bsl{\xi}_a,\bsl{\xi}_b$ average is evaluated according to the rule 
(\ref{3.9a}). Using the $\tau$ periodicity of the function $\mathcal{Q}(k,\tau)$,
the square bracket in (\ref{C.3}) becomes
\begin{align}
&\delta_{\mu\epsilon}\delta_{\nu\delta}\left[
\int_0^1 d\tau\mathcal{Q} \left(\frac{q_1}{r},\tau\right)\mathcal{Q} \left(\frac{q_2}{r},\tau\right)-\int_0^1 d\tau\mathcal{Q} \left(\frac{q_1}{r},\tau\right)\int_0^1d\tau\mathcal{Q} \left(\frac{q_2}{r},\tau\right)\right]
\nonumber\\&
=\delta_{\mu\epsilon}\delta_{\nu\delta}
\left[\frac{(\lambda_{\rm ph}/r)^2q_1 q_2}{2(1-e^{-(\lambda_{\rm ph}/r)q_1})(1-e^{-(\lambda_{\rm ph}/r)q_2})}\right.
\nonumber\\ &\left.
\quad\quad\quad\quad\quad\quad\times\(\frac{1-e^{-(\lambda_{\rm ph}/r)(q_1+q_2)}}{(\lambda_{\rm ph}/r)(q_1+q_2)}-\frac{e^{-(\lambda_{\rm ph}/r)q_1}-e^{-(\lambda_{\rm ph}/r)q_2}}{(\lambda_{\rm ph}/r)(q_1-q_2)}\)-1\right]
\nonumber\\&
\sim\delta_{\mu\epsilon}\delta_{\nu\delta}\frac{\lambda_{\rm ph}}{r}\frac{q_1q_2}{2(q_1+q_2)},\quad \frac{\lambda_{\rm ph}}{r}\to\infty
\la{C.4}
\end{align}
as shown by an explicit calculation of the $\tau$ integrals.
Inserting (\ref{C.4}) in (\ref{C.3}) and performing the vector sums 
leads to
\begin{align}
F\left(\frac{\lambda_{\rm ph}}{r}\right)\sim A\frac{\lambda_{\rm ph}}{r}
\la{C.5}
\end{align}
with
\begin{align}
A\!=\!\int \!\!\frac{d\bsl{q}_{1}}{(2\pi)^3}\int\!\!\frac{d\bsl{q}_{2}}{(2\pi)^3}\,\,
e^{i(\bsl{q}_{1}+\bsl{q}_{2})\cdot\hat{\bsl{r}}}\,\,
\frac{(4\pi)^2}{q_{1}q_{2}}\(\frac{(\bsl{q}_1\cdot\bsl{q}_2)^2}{q_1^2q_2^2}-3\)\frac{1}{2(q_1+q_2)}\;<\infty.
\la{C.6}
\end{align}
Introducing the representation $1/(q_1+q_2)=\int_0^\infty dt e^{-t(q_1+q_2)}$, the $\bsl{q}_{1}$ and $\bsl{q}_{2}$ integrals can be performed independently and each of them behaves as $t^{-2}$ as $t\to\infty$, assuring the convergence of the $t$-integral. 
Since $\left\langle \mathcal{W}_c^2(\F_a,\F_b)\right\rangle_{\b\xi_a,\b\xi_b}\sim \lambda_a^2\lambda_b^2/120r^6$ as $\lambda_a/r,\lambda_b/r\to 0$, one obtains (\ref{8.16}).


\end{document}